\documentclass[12pt]{article}

\usepackage{amsmath,amssymb,setspace,graphicx,hyperref,mathrsfs}
\usepackage{caption,subfig,comment,multicol,enumitem}

\begin{document} \title{DNNs, Dataset Statistics,\\ and Correlation Functions\thanks{Thanks to Mel Andrews, Doug Blue, Kathleen Creel, Pranab Das, Conny Knieling, Sameera Singh, Porter Williams, Stephan Wojtowytsch  for discussions and comments.}\\ \author{Robert W.\ Batterman\footnote{rbatterm@pitt.edu}\\  and James F.\ Woodward\footnote{jfw@pitt.edu}} } \date{}\maketitle 
\tableofcontents

\section{Introduction\label{intro}}

The question of how Deep Neural Networks (DNNs) work is pressing. In the philosophical literature it is often pointed out that they appear to be black boxes. This so-called ``opacity'' makes it difficult to understand their obvious and storied successes. \cite{duede_ideal_model,sullivan_Understand_ML_models,sullivan_models-represent,creel_transparency,Boge_opacity}  One way machine learners and others have attempted to precisify this question is to ask how DNNs can possibly \textit{generalize} as well as they do.  Why, that is, when appropriately trained on a given set of data, do they perform so well on new, unseen, test data?   There seems to be little consensus in the literature---either in computer science or elsewhere---about  how to answer this question. Much of the current discussion centers around why DNNs do not \textit{overfit}, as a great deal of  conventional statistics  and statistical learning theory (SLT) in the form developed by Vapnik \cite{vapnik}, apparently suggests  they should. Such overfitting will, on these views, impede DNNs ability to generalize. The overfitting problem arises because DNNs have enormous numbers of tunable parameters (often many more than the data points upon which they are trained).  This suggests that they should fit features that are noise, or idiosyncratic to that particular (training) dataset, and, consequently, should not support successful generalization to the test set.

 However, in many cases this is not what happens.  DNNs generalize successfully to the test set and often  do not overfit.  In fact, as noted in more detail below, although overfitting is sometimes observed, adding more parameters to a neural network (well beyond the point in which there are more parameters than data points) can \textit{improve} performance on the training set. In recent years, there have been a large number of papers discussing this apparent paradox (sometimes called the problem of ``double descent'' \cite{belkin_bias_var_tradeoff}) with various proposals on how it might be resolved and on what underlies the ability of DNNs to generalize successfully.  \cite{zhang_gen, rethinking_bias, Gen_in_DL,belkin_bias_var_tradeoff}
 
 These difficulties in understanding why overfitting does not occur when it apparently should be expected, are highlighted in the analysis provided by classic Vapnik-style SLT\footnote{In what follows our discussion of  SLT focuses just on what we are calling classic or ``Vapnik style SLT'' \cite{vapnik} and is confined to the possible use of this approach to understand the behavior of DNNs. We distinguish this from the general program of structural risk minimization introduced by Vapnik and its implications for various other forms of machine learning. Our discussion is not intended to cover the latter.  In this respect, we follow  recent discussions of SLT in the literature, which, as outlined above, focus on the fact that the apparent capacity of the function class that DNNs implement is enormous. On the generally accepted understanding of this ``classic'' version of SLT, such a large capacity yields trivial or uninformative bounds on the expected error for generalization. This seems to leave us with no real explanation of why there is successful generalization.  (For influential papers that understand SLT and its significance in this way, see \cite{belkin_bias_var_tradeoff} and \cite{zhang_gen}.) Also, although it is a terminological matter exactly what should be included in the general category of SLT, our discussion at this point is not intended to cover recent developments such as the literature on fast rates.  We comment briefly on this literature in section~\ref{conclusion} below. } which considers the general problem of finding bounds on the expected error on the test set  for functions selected on the basis of a training set. In particular, when applied to DNNs, this approach considers a function $f({\bf x}) = {\bf y}$ where ${\bf x}=\{x_i\}$ are the input pixelated images and ${\bf y}=\{y_i\}$ are category labels assigned to the images (cats, dogs, chairs \ldots). The DNN is given information about the correct labels for images in the training set and we are interested in the bounds on the expected error that the DNN will make in assigning labels to images in the test set, given its performance on the training set and other assumptions described below.  Error/risk is measured by a loss function that captures the seriousness of mistakes---a simple possibility is just the mean squared error. It is assumed that both the training and test data are drawn i.i.d from the same unknown joint probability distribution ${\bf P}$ governing ${\bf x}$ and ${\bf y}$. 
 
 Importantly, in the classical SLT framework we are considering, there are no additional restrictions on ${\bf P}$---it can be arbitrarily complex. It is further assumed that there is a class of functions $\mathcal{F}$ from which the fitting function $f$ is drawn. When a DNN has a very large number of parameters, it is generally thought, on both theoretical and empirical grounds, that the class of functions  $\mathcal{F}$  that it can use to fit the data is very large---large enough so that the DNN can \textit{exactly} fit the training data. 
 Slightly more technically, this is expressed in terms of the \textit{capacity} of the class of functions 
 $\mathcal{F}$.  Roughly, to say that $\mathcal{F}$'s capacity is very large is to express how flexible $\mathcal{F}$ is, in the sense that it contains some function that will fit the data regardless of what that data is.\footnote{There are different ways of measuring capacity but one standard way is by means of the Vapnik-Chervonenkis (VC) dimension and its relation to ``shattering.'' This is much more subtle than the measures of complexity and/or liability to overfitting that are standardly discussed in philosophy, such as counts of free parameters as in the Aikaike Information Criterion, and Popperian ideas about degrees of falsifiability.  There are alternative measures of capacity, such as Rademacher complexity, that are in some respects superior to the use of VC dimension but (i) these are also not fully satisfactory in accounting for DNN behavior \cite{belkin_bias_var_tradeoff} and (ii) their superiority has to do with the fact that they are more data dependent, thus reinforcing the overall message of this essay about the importance to  data structure.} The version of SLT under discussion takes the bounds on the expected error associated with generalization to be a function of the empirical risk (that is, the extent to which the DNN or some hypothesis of interest fits the training set), the size of the sample from the training set, and the capacity of the function class. When the capacity is very large, the standard analysis in terms of SLT yields bounds on the risk, given the test data, that are very large or perhaps ill-defined. Of course this leaves open the possibility that there may be other approaches, involving other sorts of restrictions, either on the function class or the data itself, that will yield better, more informative bounds on the risk---we say more about this in the text below and in footnotes. However, for present purposes, the point is that this version of SLT provides no insight into (or explanation for) why overfitting and poor generalization to the test set does not occur.\footnote{A similar point holds for other classical statistical ideas such as the claim that there is an inevitable bias/variance tradeoff. Contrary to this expectation, DNNs with apparently low bias can achieve very low variance on test data. }

%
Put very abstractly, our analysis argues that what is missing has to do with  assumptions (or rather a lack of assumptions) classical SLT makes about the data. In particular, as noted above, SLT assumes that the data from which learning occurs can conform to any arbitrary probability function---there are no restrictions on  ${\bf P}$.  Furthermore, the SLT analysis is a worst-case analysis in the sense that it provides expected error bounds that allow for the possibility that  ${\bf P}$ may be highly pathological and unfriendly to learning.

The present essay considers the  question of generalization primarily  in the context of image classification---correctly identifying previously unseen handwritten digits from the MNIST dataset or sorting images from CIFAR (and other datasets) into appropriate classes, such as dogs, cats, trucks \ldots.
We argue that the real world images on which DNNs successfully generalize, conform to very specific, non-arbitrary probability distributions.  In other words, rather than trying to locate the basis for successful generalization (solely) in restrictions on the function class the DNN is able to implement, we suggest  that the structure of the data is crucial.
Images are structured in particular ways that are friendly to learning by DNNs (and by humans too, of course). Furthermore, we hold that any account of how successful generalization is possible must take account of that structure.\footnote{For example, in their discussion of double descent, Belkin et al. \cite{belkin_bias_var_tradeoff} suggest that fit improves beyond the interpolation threshold because increasing the number of parameters allows for approximation with increasingly lower norm functions and these improve fit. \cite[p. 15850]{belkin_bias_var_tradeoff}. This strikes us as plausible as does the common suggestion that Stochastic Gradient Descent implicitly implements a preference for low norm functions (regularization). However, this does \textit{not} explain why a preference for low norm functions ``works'' in the sense of selecting functions that generalize well. We think that the answer to this  question has to do with the nature of the data that characterize images and other classificatory tasks on which DNNs succeed. Specifically, as explained in sections \ref{obj_scale} and \ref{conclusion} images themselves satisfy \textit{smoothness} constraints---pixel luminance typically changes slowly with distance---and this makes smoothness (in the sense of low norm functions) appropriate for characterizing their structure.}  

These claims raise several issues. First, what are the structural features of  images  that make  learning by DNNs possible?  We will say more about this below but they include the following, interrelated features.
We show that the actual datasets used for training possess complex, higher order non-Gaussian correlations\footnote{These are relationships that require for their characterization higher moments of a probability distribution beyond means and variances. They track  correlations among $N$-tuples of pixels for $N>2$.}  among pixels.  Such data, like visual images generally, exhibit a statistical structure (again relations among pixels) that is scale invariant (the statistics look the same at different resolutions):  They  exhibit  power law distributions.  In general, the pixels characterizing particular objects in images tend to be highly correlated with each other and  their distribution tends to be well described by smooth functions---smooth not just in the sense of differentiable but in the sense that the functions in question do not involve large changes in value within object boundaries. By contrast, object boundaries and transitions between objects involve much sharper transitions and discontinuities. This mixture of  distributions that do not involve large changes punctuated by abrupt changes is well approximated by concatenations of low-norm functions which current DNNs automatically implement through stochastic gradient descent and other features.  Yet another relevant feature is that many of the images in training sets (and of interest to humans) are relatively unambiguous---we encounter relatively few images that are intermediate between cats and dogs. Put slightly more technically, most of the data is not near the decision boundary between cats and dogs and this makes learning easier.  Our general claim, then,  is that DNNs are well-tuned  to detecting the  features just described  and using them in generalization.\footnote{How are these various features of the data related? They are connected but somewhat independent.  What we are calling the smoothness features of visual images (smoothness punctuated by discontinuities at object boundaries) does not by itself imply that the images exhibit scale invariance or conform to a power law. The latter requires that such smoothness/discontinuities be present at all scales. In other words, it is  the scale-free distribution of  smoothness punctuated by edge  structures that yields a power law. }

Taken most generally, our suggestion that the data matters may seem completely obvious. But our proposal is much more specific than this. We point to very specific features of the data that matter for image classification and we show that DNNs make use of those features.   One implication of our view is that when DNNs with current architectures are used on data that lacks the structures just described, learning may be unsuccessful. We observe below that this is, in fact, the case. 

A second issue concerns what exactly is meant by the claim that the data
  matters. A suggestion one sometimes hears is that
any specific feature of the data that matters to learning can always be re-conceptualized as a restriction on the function class available to the DNN
or other learner. For example, a referee, commenting on an earlier version of this paper, suggests that if the architecture of a DNN implements a preference for low-norm functions that fit well
with image data, then this can be regarded as a restriction on the
``effective'' function class that the DNN is able to implement, as opposed to a restriction that shows that the data matter. We respond to this objection in section~\ref{conclusion}  below.  

Overall, this paper examines the nature and genesis of the correlational structure in the actual datasets upon which DNNs are trained. In doing this it makes connections with a widespread methodology in condensed matter physics and materials science that aims to determine bulk behaviors of many-body systems (like fluids and gases) by focusing on mesoscale correlation structures that live in between fundamental, molecular or atomic scales, and continuum everyday scales.  We hope to motivate that idea that DNNs (at least in image recognition, but likely more generally) can best be understood as implementing something akin to this multi-scale methodology.  Specifically, we suggest that DNNs must be discovering high order ($>2$)-point correlation functions.\footnote{Further study of the attention mechanism and the architecture of transformer models appears to confirm this more general conclusion. However, space limitations prevent any detailed discussion in this paper.}
\subsection{Disclaimers}

Before developing our argument, let us distinguish our problem of interest---why DNNs often can successfully generalize from training sets to a test set---from issues having to do with so-called ``explainable'' or ``interpretable'' AI (XAI).  The latter has to do with approaches that seek to make the processes by which DNNs and other AI systems move from inputs to outputs understandable to, or interpretable by, humans.  These include attempts to identify the variables (where this means variables that are recognizable or interpretable by humans) that  contribute most to a classification or prediction, the construction of saliency maps which attempt to identify which regions of an image have the most influence on a DNNs classification, as well as so-called ``counterfactual explanations'' which attempt to identify those features of an input that would need to change so as to  produce a different output. These issues, as well as issues having to do with whether  and in what sense DNNs and similar systems can be said to yield understanding,  have figured centrally in recent philosophical discussions. 

 By contrast, the issue with which we are concerned---why DNNs successfully generalize and do not overfit---is not, at least not directly---an issue about whether or how the behavior of DNNs might be interpretable by humans. Although  issues about generalizability have occasioned extensive discussion in the literature outside of philosophy, they have received virtually no philosophical attention. This is true  despite the fact that these questions connect closely debates about overfitting and over-parameterization that have been  important topics both in the philosophy of statistics and the  theory confirmation over many decades.\footnote{See section~\ref{conclusion} below.} One of our goals in this essay is to draw the attention of philosophers to this problem in the context of DNNs, as well as to advance some proposals regarding its solution.  

 Our claim that the higher order correlation functions, scale invariance, and other features of data exploited by DNNs  when they successfully generalize, is consistent with the possibility that these are sufficiently complex and gerrymandered that they have no direct translation into categories and variables recognizable by humans. For this reason, we regard the explainability issue as an empirical question that is not settled by anything we say in this essay.  If this opacity to humans turns out to be the case, our claims about why DNNs successfully generalize may not tell us much about how to achieve  XAI. This is not to say that some variety of XAI is impossible, but rather that if it is possible, it will need to focus on questions distinct from the question of how DNNs work as well as they do.\footnote{One might think (probably some philosophers do think) that if  (i) explainable AI is achieved, this will automatically answer the question  (ii)  why DNNs work as well as they do. We think this is far from obvious. For one thing, note that the answers given to (ii) both in this essay, and elsewhere in the computer science literature, that emphasize the role of higher order correlation functions, scaling relations, etc.\  are very different from those emphasized in the  literature on explainable AI. In the latter, interpretability in terms of variables  and relationships that are understandable by humans, and that are relatively macro-level (as with saliency maps or answers to counterfactual questions about altering inputs)  are key.
However,  the features of DNNs that best explain why they work need not be constrained in this  human-centric way. Likewise, our focus, in this essay and in the literature on how DNNs generalize as well as they do, concerns generic features of DNNs that enable the learning of higher order correlation functions etc. This is very different from explaining why some particular DNN, with a given architecture and trained in a certain way, works in the sense of making certain classification decisions rather than others. That is typically the concern of XAI.
Finally, we note that  the project of altering current  DNNs so that their behavior is  less opaque, more interpretable, or more readily controllable  by humans makes perfect sense.  To say this is, of course, to agree that current DNNs may not be very explainable in the sense sought by XAI. However, it does not mean  that there is no explanation of the sort that we are trying to give for why DNNs work as well as they do.}
 
 A similar point holds about current philosophical discussions regarding whether DNNs (and LLMs etc.) exhibit ``understanding.'' ``Understanding'' is a very vague word and we take no stand on this question, except to remark that a better grasp of how DNNs work and successfully generalize might reasonably be expected to provide some guidance on this issue. 
 

\subsection{Outline}

In the next section we report on some pioneering work on image statistics from the 1990s that explicitly takes a correlation function approach to understanding robust statistical features in datasets. Humans, in fact, use these statistical features in learning to segment visual scenes into distinct objects. We suggest that DNNs likely do the same thing. Object segmentation is different than object classification---determining that a particular image is of a specific kind (dog vs.\ cat). We provide further evidence (section~\ref{implement}) that classification tasks require appeal to higher order correlations. This use of statistics in data segmentation is further evidence of our general theme that worldly facts about the data structures matter---in this case, the empirical fact that pixels similar in their luminance are likely to belong to the same object.

In section~\ref{cf_method} we briefly elaborate on the correlation function methodology  mentioned above. This is followed in section~\ref{dataset}  by a more detailed discussion of the correlational statistics found in the actual datasets used in training and testing and further  connections between those statistics and the evolution of the statistics of the layer weights in real DNNs as they are trained on those datasets.  Section~\ref{implement}  
addresses two important questions:  (1) Are higher order correlation functions sufficient to distinguish members of one class (say, cats) from another (say, dogs) in the same dataset?  We provide evidence that this is indeed the case. (2) Given a positive answer to the first question, are DNNs \textit{actually} finding such higher order correlation functions? Here we discuss some recent work that suggests that this question should, as well, receive a positive answer. The discussion here makes connections with certain perturbative calculations in quantum field theory that enable the approximation of $N$-point correlation functions (Green's functions). In so doing it further supports our contention that DNNs are implementing the multi-scale methodology discussed in section~\ref{implement}.

\section{Natural Images: Objects and Scaling\label{obj_scale}}

\begin{figure}[h]
\centering
    \includegraphics[width=.45\textwidth]{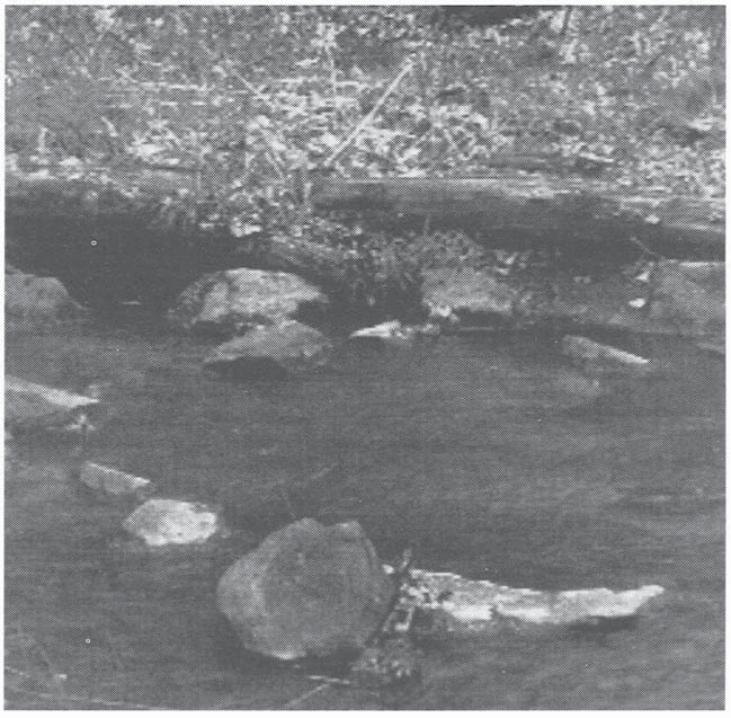}\caption{Stream, Trees, Rocks}
\label{woods}
\end{figure}

Ruderman and Bialek \cite{ruderman-Bialek} took a series of photographs at Hacklebarney State Park in New Jersey. The photos were primarily of trees, rocks, and a stream. An example is displayed in figure~\ref{woods}. The images measured 256 by 256 pixels and corresponded to 15 degrees in visual angle. The data they collected were the logarithm of each pixel’s luminance.\footnote{\label{freq}The pixel luminance is, essentially, the value of the pixel's brightness. (The luminance has  been weighted according to how humans perceive that brightness.) The  (spatial) frequency $k =(k_x,k_y)$, in equation~(\ref{scaling}) is a two dimensional vector representing how rapidly the intensity oscillates in  the horizontal and vertical directions. }  \cite[p. 3386]{ruderman1997}. 
The data showed scaling ``in the power spectrum of the form\footnote{The power spectrum is defined as the Fourier transform of the (auto)correlation function between the log contrast at pixels ${\bf x}$ and  ${\bf x} + {\bf y}$: $S_\phi{(\bf k}) = \int e^{i{\bf k} \cdot {\bf y}} \langle\phi({\bf x})  \phi({(\bf x} + {\bf y})\rangle$, where the average, $\langle \cdots \rangle$, is over the images. \cite[p. 814]{ruderman-Bialek}}

\begin{equation}\label{scaling}
S(k) = \frac{A}{k^{2-\eta},} 
\end{equation}
with $k$ being the spatial frequency, $A$ is a constant representing the overall contrast power in the images \ldots.'' The power spectrum essentially captures how much fluctuation in contrast the images in their collection have as a function of spatial frequency.\footnote{An image with large structures like tree trunks but with \textit{few} fine details like leaf structure will have its power $S(k)$ concentrated in low frequency ranges of $k$. One with \textit{more} detail would have the power concentrated at higher frequency ranges.}
 
 For their data the “anomalous” exponent $\eta$ had a value of 0.19. This means that the images in the dataset exhibit structures that are not seen in Gaussian distributions (had the exponent been equal to 2). 

\begin{figure}[h]\centering
\includegraphics[width=.7\textwidth]{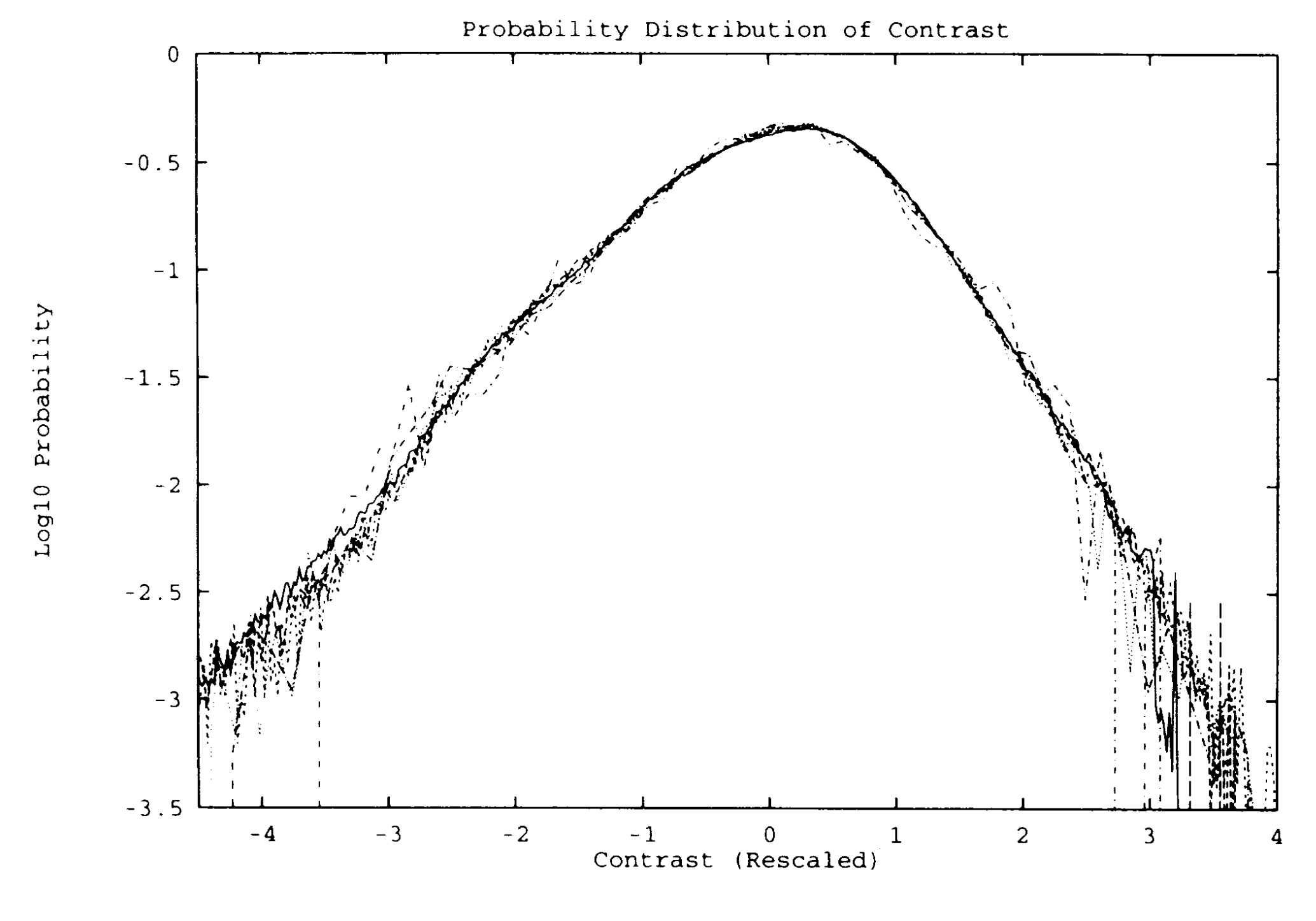}
\caption{\label{contrast} \textit{Universal} Scaling of Contrast Distribution. Plots for each of $N^2$ pixel blocks, ($N=1, 2, 4, \ldots, 32$)}
\end{figure}

This scaling result means, essentially, that if one zooms in or out by changing the focal length of one's camera lens, the resulting images will statistically be the same. Such changes (e.g.  zooming out) will be equivalent to a coarse-graining or averaging operation. Close-up pictures have more pixel detail. Zoomed-out pictures provide  a kind of average over that detail which nevertheless continue to be (recognizable) pictures of the same real-world scene.  More technically, Ruderman and Bialek show, in Figure~\ref{contrast}, that the probability distribution of the contrast $\phi$, of the images (normalized to unit variance) averaged over $N^2$ pixel blocks for ($N=1, 2, 4, \ldots, 32$) are all identical.\footnote{The contrast of pixel $\bf x$ is defined as $\phi({\bf x}) = ln [I({\bf x})/I_0]$, where $I({\bf x})$ is the intensity of ${\bf x}$ and where  $I_0$ is a normalization factor for each image that ensures that the average contrast over all images in the dataset is zero. Note that Figure~\ref{contrast} superposes \textit{all} of the (non-Gaussian) plots for ($N=1, 2, 4, \ldots, 32$). \cite[p. 815]{ruderman-Bialek}}  The different pixel blocks $(1\times 1)$, $(2 \times 2)$, \ldots implement the averaging/coarse-graining (equivalent to zooming out). Figure~\ref{contrast} shows that such an operation leave the distribution of contrast throughout the image unchanged.  The \textit{shapes} of the distinct plots for the $N\times N$  blocked images seen in Figure~\ref{contrast} are all the same. This fact is ultimately responsible for the scaling law, equation~(\ref{scaling}). This means that all the plots are instances of the same (universal) $1/k^{2-\eta}$ scaling. Importantly, the plots are non-Gaussian exhibiting so-called fat tails.\footnote{Fat-tailed distributions fall off more gradually (as one moves away from the mean) than do Gaussians which decay exponentially. This means, as we show below, that the images possess correlational structure absent in Gaussian/normal distributions.}
In fact, this result is remarkably robust:
\begin{quote}
That the process of geological formation of hillsides and valleys, or the structure of forests due to the succession of ﬂora, can exhibit scaling through their images is perhaps not altogether surprising. . . . It is striking, however, that the natural image datasets in which scaling was found are all quite different. No two sets of pictures were even from the same environment. \cite[ pp. 3385-3386]{ruderman1997}
\end{quote}

Another indication of the robustness of the statistical structure in the natural images is shown by implementing a rather radical recalibration of the data. Ruderman describes a simple experiment in which all of the gray scale images in the data set were converted to black and white.\footnote{If the logarithm of a pixel’s luminance was greater than zero it becomes white, otherwise it becomes black. \cite[p. 3389]{ruderman1997}} This produced a new dataset, yet the (non-Gaussian) statistics---the power law exponent---is virtually unchanged:  The exponent $\eta$ ``changed slightly from $\eta  = 0.19$ to $\eta' = 0.20$. Given the drastic nature of the recalibration procedure, this change is surprisingly small.'' \cite[p. 3389]{ruderman1997}

This example is meant to demonstrate the robustness of scaling in natural images by pushing an extreme limit of recalibration. Of course it cannot be expected that entirely arbitrary recalibrations (e.g., a random reassignment of pixel values) would preserve the correlational structure of the images. More reasonably, it probably holds as long as nearby pixel values generally remain nearby under recalibration. \cite[p. 3389]{ruderman1997}

The scaling result (\ref{scaling}) describes a relation in the \textit{frequency} domain---it is a function of the  frequency $k$. (See footnote \ref{freq}.) As Ruderman notes, analyzing images in the frequency domain is not the best way to understand which properties of natural images are responsible for scaling.\footnote{The frequency domain, essentially, looks at how much variance exists at each spatial frequency, independently of  location within the image. This lets one easily define the power spectrum in equation~\ref{scaling}.  The spatial domain, on the other hand,  looks at the distribution of pixel intensities (luminance) and how they differ from neighboring pixels.} This is  because we actually observe images in the spatial domain: ``Objects, after all, are generally spatially cohesive. In the Fourier domain, though, they spread and superpose over many frequency bands.'' \cite[p. 3387]{ruderman1997}. So Ruderman reformulates the results in the spatial domain, introducing correlation functions that allow him to ``define'' objects \textit{statistically}.

He introduces a correlation function that gives the expected product of the data at two pixels separated by a distances $x$:

\begin{equation}\label{corr}
C(x) = \langle\langle\langle\phi({\bf x_0})\phi({\bf x_0}+{\bf x})\rangle_\theta\rangle_{{\bf x_0}}\rangle_\phi.
\end{equation}

\noindent $\phi({\bf x})$ is the image value\footnote{Again, this is the logarithm of the luminance (related to the brightness or intensity).} at position ${\bf x}$ and the expectations (from the outside in) are taken over all the images $\phi$, all initial positions ${\bf x_0}$, and all displacement vectors ${\bf x}$ of length $x$ parameterized by the angle $\theta$. \cite[p. 3387]{ruderman1997} Ruderman next introduces a ``difference function'' which is linearly related to $C(x)$:
\begin{equation}\label{diff_fun}
 D(x) = \mathop{\pmb{\langle}}|\phi(0) - \phi(x)|^2 \mathop{\pmb{\rangle}},
\end{equation}
where the bold angle brackets stand for the three expectations as in (\ref{corr})).\footnote{The difference function is a kind of expected variance. If its value is small, it is likely that the pixels belong to the same object.}  This allows one to easily define objects in images in terms of probability distributions: There will be a probability $P_{\mbox{\tiny{SAME}}}$ that a given pair of pixels a distance $x$ apart  belong to the same object. 

Now consider a model of image generation that puts ``randomly chosen objects in the world at random locations'' allowing objects to occlude one another. \cite[p. 3389]{ruderman1997} One illuminates the world and takes a picture. Given this model, since objects are chosen at random, pixels in the images that correspond to the same object will have greater statistical dependence on each other than those from different objects, in part because ``of the likelihood of them originating from the same material and receiving similar lighting.'' \cite[p. 3389]{ruderman1997} Choose two pixels at random and calculate the value of $D(x)$ for degree of visual angle $x$.\footnote{The degree of visual angle is  proxy for the spatial separation between the pixels. It is the angle subtended from the camera's lens between those two points.} ``[T]his probability depends on the actual spatial sizes of objects, their distribution of distances from the observer [at $x = 0$], and their shapes.'' \cite[p. 3389]{ruderman1997}   

For two pixels belonging to the same object separated by $x$, there will be a corresponding difference function, $D_{\mbox{\tiny{SAME}}}$. Likewise for two pixels belonging to different objects, there is a difference function $D_{\mbox{\tiny{DIFF}}}$.\footnote{\label{DdDs}$D(x)$ is determined by averaging the squares of the pixel differences between any two points in an image.  \cite[p. 3389]{ruderman1997}  $D_{\mbox{\tiny{SAME}}}(x)$ averages the squares of the pixel differences between two points from the same object in an image.  Analogously, $D_{\mbox{\tiny{DIFF}}}(x)$ does the averages of squares for points from different objects in an image.} We can then express the distance function $D(x)$ as follows:

\begin{equation}
D(x) = P_{\mbox{\tiny{SAME}}}(x) D_{\mbox{\tiny{SAME}}}(x) + [1 - P_{\mbox{\tiny{SAME}}}(x)] D_{\mbox{\tiny{DIFF}}}(x).
\end{equation}
By examining the images in the ensemble, making reasonable assumptions about what counts as an object in an image by appealing to “gross semantic boundaries,” Ruderman is able to determine values for $D_{\mbox{\tiny{SAME}}}$ and $D_{\mbox{\tiny{DIFF}}}$.\footnote{He says, ``[f]or example, [the image shown in the paper] was divided into regions corresponding to the stream, the rocks, the riverbank, the log on the river, etc. \ldots leaves on trees were considered integral parts instead of objects in their own right \ldots . Suﬃce it to say there there is no entirely objective way of doing this.''   \cite[p. 3389]{ruderman1997}} 

We mentioned, above, Ruderman’s model of image generation. It is worthwhile going into a bit more detail about it. It is introduced as follows:

\begin{quote}Imagine walking on an inﬁnite image plane. At a random location you blindly select from a number of choices an inﬁnitesimally thin cardboard ``cut-out'' of some shape. You paint it a gray tone chosen from a distribution, and then drop it on the ground. This done, you continue to another random location and repeat the process. \cite[ p. 3392]{ruderman1997}
\end{quote}
Such a model involves statistically independent objects that can occlude one another. ``The true `independent components' are the objects themselves, which have random size, location, and intensity.'' \cite[ p. 3392]{ruderman1997}

 Ruderman establishes two sufficient conditions for the scaling of correlations within the images. The ﬁrst is that the probability distribution of ``not crossing an object border scale in distance.'' The second is that ``objects have nearly uniform correlation within their borders [and zero correlation] between different objects.''\footnote{There is a typographical error in the paper that omits something like the bracketed phrase in this sentence.} \cite[ p. 3392]{ruderman1997}. This second sufficient condition is guaranteed in the model by the fact that the cardboard objects are randomly painted with distinct gray tones before being placed.
 
As this will be relevant for the later discussion (section~\ref{smooth}), let us say a little more about this model and its relevance for   the relationship between the scale invariance (the observed power law behavior) and the ``smoothness'' one finds in natural images:  Within an ``object'' (the cutout of a given shape) there will be little or no variation in the  luminance of different pixels.\footnote{In the model, there is no variance by design; in real world images like that of figure~\ref{woods}, the variance will be small between pixels belonging, e.g., to the same tree trunk.} This is the \textit{smoothness} that can be exploited by DNNs. The fact that the distinct objects can partially occlude one another, means that there will be large jumps in luminance between the different ``objects.'' These jumps are reflected in the values for $D_{\mbox{\tiny{SAME}}}$ and $D_{\mbox{\tiny{DIFF}}}$. (See footnote \ref{DdDs}). The power law scaling in equation~(\ref{scaling}) results from the fact that relative ratio between smooth (object) regions to disjoint transitions (at edges) is self-similar: Whether one zooms in or out, one will see the same ratio.

This model has a correlation function $C(x)$ deﬁned as follows:

\begin{equation}
C(x) - C_0 P_{\mbox{\tiny{SAME}}}(x),
\end{equation}
where $C_0$ ``is the constant correlation within objects, and the term for different objects is absent since they have zero correlation.'' \cite[ p. 3392]{ruderman1997}. It remains to determine the value of $P_{\mbox{\tiny{SAME}}}(x)$. If this has power law scaling then so does the correlation function $C(x)$. Ruderman demonstrates that $P_{\mbox{\tiny{SAME}}}(x)$ is indeed a power law.

\begin{figure}[h]\centering
\includegraphics[height=.5\textheight,]{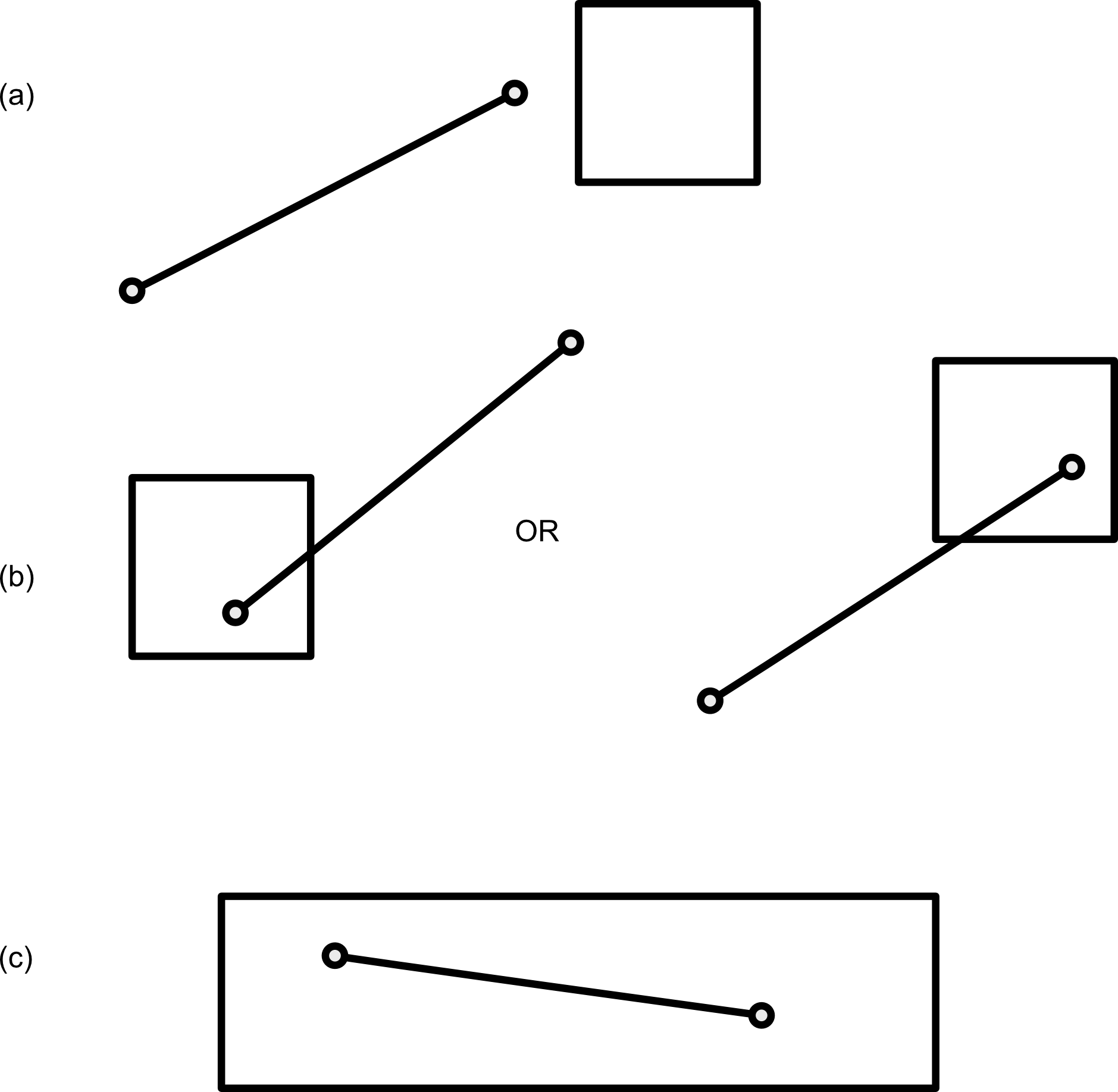}
\caption{\label{segments} Calculating 2-point Correlations. Rectangles are Regions/Objects. \cite[p. 3393]{ruderman1997} (a) yields $p_0(x)$; (b) yields $p_1(x)$; (c) yields $p_2(x)$ .}
\end{figure}

Without going into the details of the calculations, it is instructive (for the discussion to come) to get a qualitative understanding of the reasoning involved. First he shows that one can rewrite the formula for $P_{\mbox{\tiny{SAME}}}(x)$ as follows:
\begin{equation}
P_{\mbox{\tiny{SAME}}}(x) = \frac{p_2(x)}{p_1(x) + p_2(x)},
\end{equation}
where $p_1(x)$ and $p_2(x)$ are determined by examining ﬁgure~\ref{segments}.\footnote{Note that $1 - p_0 = p_1 + p_2$, so $p_0$ (the probability that neither pixel separated by length $x$ is in the object) drops out of the formula for $P_{\mbox{\tiny{SAME}}}(x)$.}

Given conﬁguration (b) in figure~\ref{segments}, $p_1(x)$ is the probability that, for a pair of pixels separated by length $x$, exactly one of them lies in a given region. Given conﬁguration (c), $p _2(x)$ is the probability that, for a pair of pixels separated by length $x$, both lie in a given region. Finally, conﬁguration (a) yields the probability that neither pixel separated by length $x$ lies in a given region.\footnote{It is an interesting historical fact that Ruderman's  2-point correlation function calculation was already performed by Debeye, et al. \cite{Debeye_cf}. See also, \cite{Chalkley_cf} for an even earlier related calculation.}

Ruderman concludes that
\begin{quote}
\ldots the scaling of inter-object probability follows directly from the scaling of apparent object sizes. In images of the real world this apparent size (in degrees) depends on an object's actual size as well as its distance from the observer. The overall distribution of apparent object size is thus a function of the distributions of object sizes and that of their distances. \cite[ p. 3393]{ruderman1997}
\end{quote}

\noindent For the purposes of our argument in this paper, the most important feature of this conclusion is the fact that scaling invariance in the images can be recovered by a model that both treats the segmented objects within images \textit{statistically} and employs the \textit{correlation function method} exhibited in Figure~\ref{segments}.  We elaborate further on this in   section~\ref{cf_method} below.

%

In this section, we have gone into quite some detail about how to describe and explain scaling in natural images. The main reason for this is to suggest that the effectiveness of DNNs  depends crucially upon there being correlational structure in the datasets input to the DNNs. In other words, our discussion so far shows that natural images exhibit scale invariance that is described by correlation functions. We take this to help motivate the arguments below that DNNs learn to classify images by learning this correlational structure.
This is to say that what can be called ``worldly structure'' in the data is an important, necessary ingredient  for understanding many of the successes of deep learning. Nevertheless, we   still need an account that shows how DNNs are able to exploit the presence of certain statistics in the data. 



The next section relates Ruderman's appeals to \textit{correlational structure}, both as  means for \textit{statistically defining}  objects in images and for determining  power law scaling exponents, to a broader scientific methodology for understanding the macro-level behaviors of many-body systems. For instance, materials like fluids exhibit different continuum scale behaviors because they have different viscosities.  Materials like copper and asbestos exhibit different continuum scale behaviors for heat conduction because they have distinct values  for heat diffusivity (or conductance).  ``Continuum scale'' or ``macro-level'' behaviors, here, refer to robust behaviors that are represented mathematically  by continuum equations such as the Navier-Stokes equations for fluids, and the heat equation for conducting materials. Such equations completely ignore structures below the macro-level and treat fluids and materials as continuous blobs. They countenance no structures at molecular, atomic, and subatomic levels. They also describe our everyday (macro) experience with such systems.

A natural question to ask concerns what determines the values for continuum parameters like viscosity and diffusivity. The explanation is to be found not at the micro-level of molecules or atoms, but rather by examining ``mesoscale'' structures---scales intermediate between the micro-level and the macro (continuum) level. That is, looking only at micro-level components is \textit{insufficient} for determining   values for macro-level or continuum parameters.  Such mesoscale structures consist of sometimes quite complicated (spatial) correlations between the more micro components.   The example in the next section~\ref{cf_method} shows that the degree to which a composite solid (of a conductor and an insulator)  is a good heat conductor depends upon the geometric and topological arrangements  of the components within the material. We can learn about such geometric and topological arrangements by examining correlation functions that characterize correlations of higher order than the 1-point (mean) and 2-point (variance) that Ruderman employs in his analysis. In other words, the statistical characterization of the arrangements will involve moments of the probability distribution describing the arrangements of the different components that go beyond these 1-point and 2-point correlation functions. 

Most importantly, as we argue below, such higher order correlation functions are necessary for image \textit{classification}. Ruderman's 1-point and 2-point correlation functions (yielding values for $p_0, p_1, p_2$ above), while sufficient for determining whether two pixels belong (or not) to the same object in an image (sufficient, that is, for object \textit{segmentation}),  are \textit{not} sufficient for determining whether an object in an image is that of a dog or of a cat. (That is, 1-point and 2-point correlations are insufficient for image \textit{identification or classification}.) As we show below, just as characterizing  mesoscale structures represented by $N$-point correlation functions ($N>2$) can explain macro-level (viscosity) differences  between materials (e.g., honey vs.\ water), 
such higher order $N$-point correlations between image pixels can lead to membership in different classes of objects (dogs vs.\ cats).



In the context of image recognition/classification, the many-body analogs of individual atoms or molecules are the   pixel inputs for different images; and, the many-body analogs of material bulk behaviors (captured by different values for the heat diffusivity or viscosity) are the labeled facts that the image is  of a dog,  of a cat, of a car \ldots . Put more metaphorically, but still accurately, the aim of DNNs (like the aim of determining values for continuum parameters such as heat diffusivity or viscosity) is to upscale from atomic (pixel) level details to distinct ``continuum level'' parameter values like $dog$, $cat$, $car$ \ldots. The key, in both cases, is to find higher order correlational structures---correlations that exist at the mesoscale between positions of individual molecules in the case of materials and between pixels in the case of images.


\section{A Correlation Function Methodology\label{cf_method}}

As just noted, Ruderman's scheme for determining the two-point correlation functions between image pixels is actually an instance of a widely applicable multi-scale methodology for understanding the behaviors of many-body systems in condensed matter physics and in materials science.\footnote{This methodology was promoted by Leo Kadanoff and Paul Martin. It has its roots in Einstein's work on Brownian Motion. See \cite{kad-martin} for the original paper and \cite{forster-hydro} for an extended discussion. See also \cite{middleway} for a philosophical discussion of the importance of this methodology and its relation to various philosophical issues concerning the relations between theories at different scales.} It is sometimes referred to as a set of ``hydrodynamic''  or ``correlation function'' methods. 

In order to characterize upper-scale/bulk behavior of such many-body systems, the most important continuum scale, i.e, \textit{macroscopic}, quantities are so-called ``material parameters'' and ``order parameters.''  Examples of these, respectively, include the viscosity of a fluid and the net magnetization of a ferromagnet.  For instance, using the Navier-Stokes equations to describe, predict, and explain the behavior of a particular fluid, requires that one  determine the values for the   density and viscosity parameters that appear in those equations.  While in most instances, one finds these values by laboratory experiments, theoretically   such  parameters are actually coding for \textit{correlational structures} at mesoscales in between  the so-called ``fundamental'' atomic or molecular scales, and continuum scales.\cite{middleway}   Such correlational structures are \textit{hidden} at atomic scales and  only become \textit{visible} at mesoscales.

For illustrative purposes, let us consider the heat equation which describes  how heat diffuses through a material:
\begin{equation}\label{heateqn}
\frac{\partial u}{\partial t} = \alpha \left(\frac{\partial^2 u}{\partial x^2} + \frac{\partial^2 u}{\partial y^2} +  \frac{\partial^2 u}{\partial z^2}\right);
\end{equation}
where $u(x,y,z,t)$ is the temperature of the material at spatial point $(x,y,z)$ at time $t$, and $\alpha$ is a parameter known as the thermal diffusivity of the material.  This equation is an  ``effective'' equation that describes the behavior of a  continuum field---the heat field. As a continuum equation it posits no structure at scales below the macroscopic. In order to actually employ equation~(\ref{heateqn}) we need to know the  diffusivity of the material, $\alpha$, a real-valued material parameter. 

\begin{figure}

\centering
\scalebox{.7}{\includegraphics[width=1\textwidth]{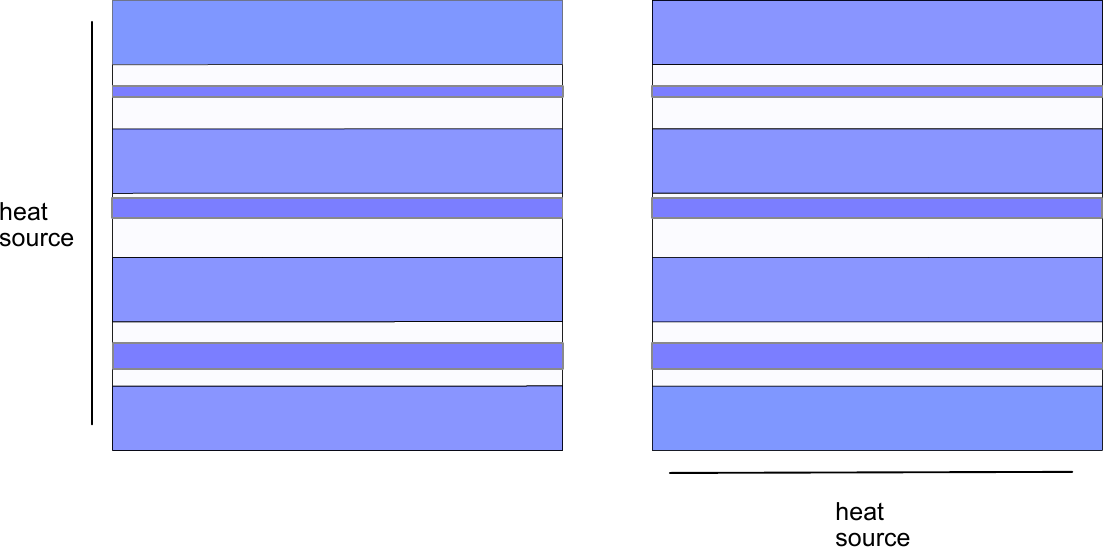}}
\caption{Conductor/Insulator Composite. Dark bands are  the Insulators.\label{condinsul}}
\end{figure}

Now consider a material that is a composite of a heat conductor and an insulator with a sandwich-like structure as in figure~\ref{condinsul}.
 At the macroscopic, or continuum, scale this sandwich structure is \textit{not} discernible. At that scale, the material appears to be completely homogeneous. We would like to determine the \emph{effective} diffusivity, $\alpha_e$, as a function of the diffusivities of the two materials. Suppose that the dark material is the insulator with $\alpha = \alpha_I$ and the light material is the conductor with $\alpha = \alpha_C$, where $\alpha_C > \alpha_I$. It is clear that if we heat up the left side of the material, then after some time $\Delta t$ the temperature on the right side will be considerably higher than the temperature would be at the top, had we instead heated the bottom with the same heat source and measured the temperature after the same elapsed time. 

In fact, the \emph{effective} value of $\alpha$ for the entire composite in the lefthand configuration is 

\begin{equation}
\alpha_e = \alpha_I\phi_I + \alpha_C\phi_C,
\end{equation}

\noindent where $\phi_I$ and  $\phi_C$ are the volume fractions of the insulating material and the conducting material respectively. This is the arithmetic average.\footnote{Note that $\phi_I +  \phi_C = 1$.} If one believed that this average was the effective value for the diffusivity, one would grossly \textit{overestimate} the heat conductivity of this example material at continuum scales. This is because the effective value,  $\alpha_e$, in the righthand configuration (where the heat source is at the bottom) is best represented by the harmonic average:

\begin{equation}
\alpha_e = \left(\frac{\phi_I}{\alpha_I} + \frac{\phi_C}{\alpha_C}\right)^{-1}.
\end{equation}

\noindent If one, likewise, believed that this (harmonic) average was the effective value for the diffusivity when the heat source is on the left, one would grossly \textit{underestimate} the heat conductivity of the material at continuum scales.\footnote{This discussion follows that of  \cite[pp.10--11]{torquato}.}
This example shows that geometric structure at the mesoscale is relevant for the conductive behavior of the material at the macroscale. Here, a mesoscale, geometric notion of structure is defined by the geometric and topological  arrangement of the component conductors and insulators. This determines the macroscopic, continuum scale, material property: The diffusivity. 
For materials that are heterogeneous at mesoscales, then, the effective values of the material parameters appearing in our continuum equations are essentially dependent upon geometric structure at scales in between the atomic and the continuum. 
The material parameters appearing in our effective equations are coding for structures at those \textit{meso}scales.

In the last section, we saw that Ruderman's line segments (and the probabilities $p_0$, $p_1$, and $p_2$) allow for the determination of two-point correlations between the pixels' luminances. This provides some information about the statistical structure of the image---information that, as we have seen, is sufficient to determine the power law scaling of the images in the dataset. 
\begin{figure}
 \centering
\includegraphics[height=.4 \textheight, trim=2in 4in 2in 3in,clip ]{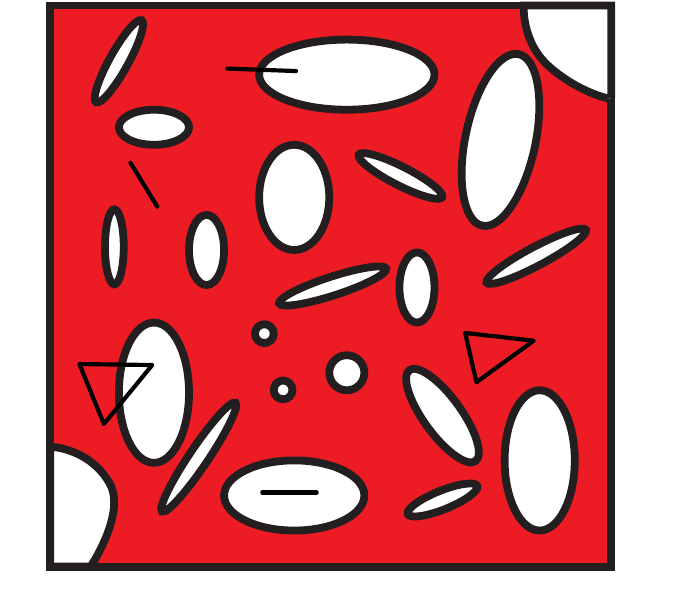}

  \caption{Using Lines, Triangles  \ldots to Determine 2-, 3-, \ldots, $N$-point Correlation Functions. \cite{middleway}\label{npoint}}
  \end{figure}
\noindent In the current context however, to determine   \textit{theoretical} values for the continuum scale effective diffusivity of the composite, one  needs considerably more correlational information.  As illustrated schematically in figure~\ref{npoint}, such information can be obtained by finding higher order (3-point, 4-point, \ldots, $N$-point) correlations. Higher order correlation functions will allow us to recognize that the light colored phase in figure~\ref{npoint} is topologically disconnected and the dark phase is connected. If, for example, the material was a semiconductor and the dark phase was an insulating material, upscaling with  information only from 1-point and 2-point correlations will grossly underestimate the conductivity of the material.  With $N$-point correlations for large $N$ we can gain information that allows for the determination of continuum scale (field value) for the conductivity/diffusivity, $\alpha_e$, of the material as $N\to\infty$. 

Unfortunately, correlation functions of order greater than two are extremely difficult to calculate. 
All sorts of approximations and  simulations have been proposed, very few of which in practice go beyond order three.\footnote{For an idea of what is involved see \cite{new_n-point_cfs}.}
In the case of the two ``sandwich'' composites of figure~\ref{condinsul}, we \textit{know} the mesoscale structure and that allows us to determine the different diffusivity values depending on the location of the heat source. In general, the actual mesoscale structure maybe  complex and unknown.
Various kinds of scattering experiments can then be used to gain certain correlational information. In other cases, materials scientists often start with what they believe to be a statistically \textit{representative volume element} (RVE). For instance, on the assumption that figure~\ref{npoint} is an RVE, they will seek to determine $N$-point correlation functions to determine estimates for continuum scale material parameters such as the diffusivity above or  Young's modulus for elastic materials.

 We have argued that the effectiveness of DNNs in image recognition (among other tasks) depends on the existence of correlational information that reflects features of the real world. Below we   argue that DNNs must be finding $N$-point correlation functions present in the input (pixel)  data. So, rather than \textit{starting} with a RVE, as is often the case in materials research, we suggest that DNNs can be understood as \textit{constructing or finding representative volume elements} (or at least the correlations they code for) for distinct classes within the various datasets. That is, we propose that DNNs are finding higher order correlation functions  that, essentially, characterize RVEs for classes like dogs, cats, trucks \ldots.

\section{Dataset Statistics\label{dataset}}

An influential paper by H. Lin, M. Tegmark, and D. Rolnick entitled ``Why Does Deep and Cheap Learning Work So Well?''\cite{lin-deep-cheap}, aims to show how the success of deep learning depends, not only on the mathematics of neural networks but also on certain facts about the world. They frame this as follows. ``How can neural networks approximate functions well in practice, when the set of possible functions is exponentially larger than the set of practically possible networks?'' \cite[p. 1225]{lin-deep-cheap} The question arises because even networks with only one hidden layer can be shown, mathematically, to be universal function approximaters. That is, given a sufficient number of hidden units any smooth function can be approximated to any accuracy with just a single hidden layer. Lin et al.\ give a quick estimate that demonstrates that networks of ``feasible size'' however cannot do this. ``There are $2^{2^n}$ different Boolean functions of $n$ variables, so a network implementing a generic function in this class requires at least $2^n$ bits to describe, i.e., more bits than there are atoms in our universe if $n > 260$.'' \cite[p. 1228]{lin-deep-cheap} Despite this, neural networks of ``feasible size\footnote{Read ``actually implementable.''}'', have been extremely successful. 

Lin et al.\  try to explain this success by noting that scientists who use neural networks only care about some small fraction of all functions that can be approximated. They argue that the kind of functions scientists/physicists typically care about are Hamiltonians of low polynomial order. Often these functions display certain symmetries and reflect local interactions. \cite[Section 2.4]{lin-deep-cheap}
These considerations help, they claim, to explain why we can get away with ``relatively'' small neural networks: The kind of functions we want to approximate are extremely far from being random. In effect, they argue that one reason DNNs work well is because the space of functions we actually care about is extremely small in the space of all functions.

We understand their argument as suggesting that worldly considerations may play a role in explaining DNNs successes at generalization.  Appeal to symmetries and locality reflect facts about the dynamics of our world. However, it is not clear that the functions DNNs actually implement are simple, familiar ones. On the contrary, if they are functions that compute or approximate $N$-point correlations, then they are likely to be quite complicated. Furthermore, even if DNNs implement simple functions that reflect our interest in physics contexts, more would be need to be said concerning why and how these functions relate to successes in the quite different context of image recognition.  The argument of the previous section suggests that DNNs are approximating $N$-point correlation functions because these are the \textit{natural} functions with which to characterize bulk behaviors in many-body systems.\footnote{There is a great deal of other evidence that many features of human visual perception rely on higher order statistics. See, e.g., \cite{portilla_et_al}  and the references therein.}

So, we believe the real interest is in how DNNs actually find the functions that \textit{work}---the functions that correctly recognize objects at the scale of dogs, given input at the scale of pixels. These are the functions that we should care about (not simply some set of functions with low polynomial order, etc.).   The explanation for this must, as noted, appeal to actual statistical facts about the world. Such facts include, but go beyone, the scale invariance Ruderman finds in natural images.  We  want to understand how DNNs find functions that detect the correlations that yield that invariance as well as  other robust statistical regularities.

\subsection{Scaling in Datasets\label{scale_data}}

So here we need to look to the datasets upon which DNNs are trained. A partial list of these datasets include:
\begin{itemize}
\item MNIST---A large database of handwritten numerals.
\item FMNIST---An MNIST-like database of labeled fashion images.
\item CIFAR10---A very large database of labeled images from 10 classes representing airplanes, birds, cars, cats, deer, dogs, frogs, horses, ships, and trucks.
\item IMAGENET---A huge database containing more than 14 million labeled images from more than 20,000 classes or categories.
\end{itemize}

\begin{figure}[h]\centering
\includegraphics[height=.45\textheight,trim=7in 23in 16.5in 15.5in,clip]{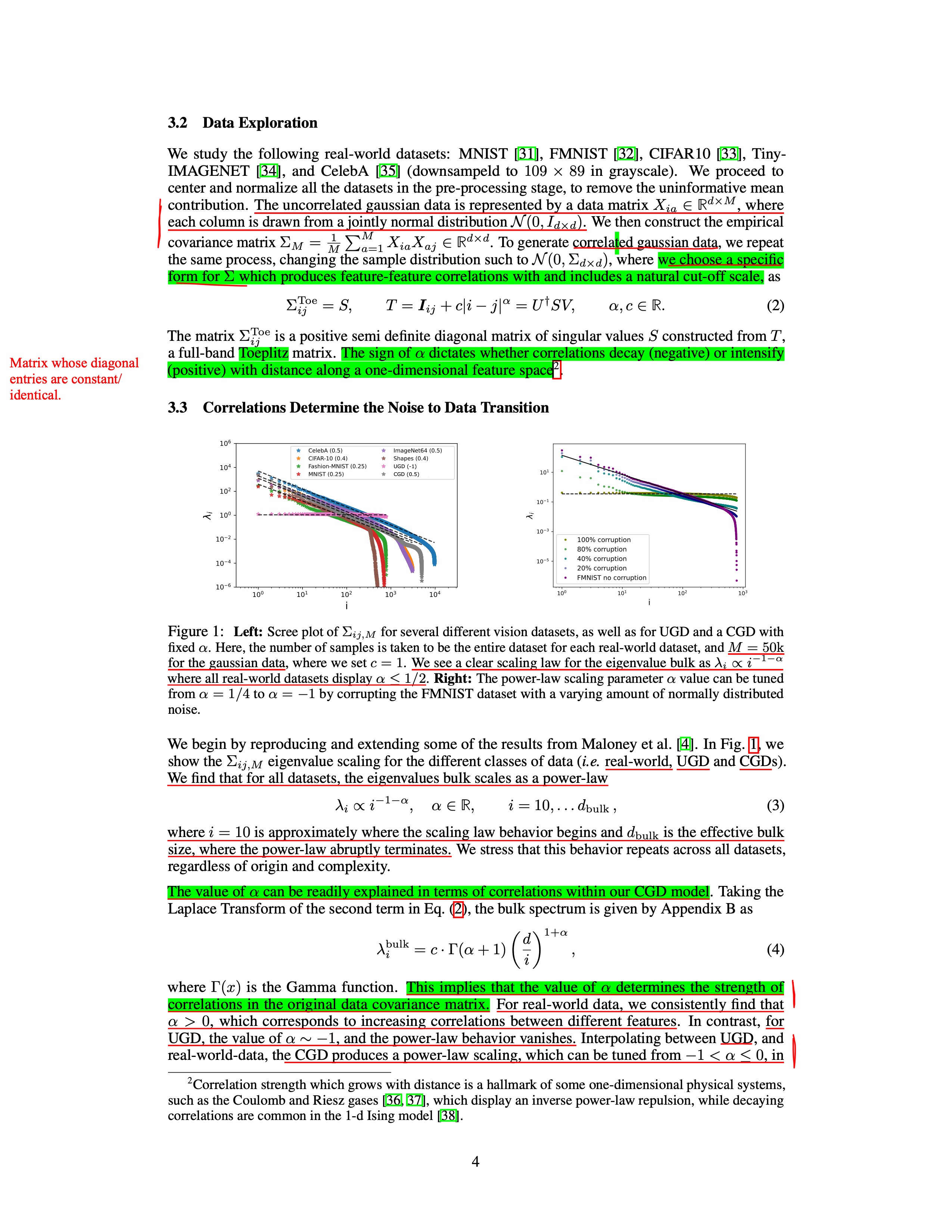}
\caption{\label{scree}Scree Plots:  Scaling Behavior of $\Sigma_M$ for Various Datasets (``CGD'' stands for Correlated Gaussian Datasets. (``UCD'' stands for Uncorrelated Gaussian Data.) The parenthetical $\alpha$'s are the anomalous exponents corresponding to Ruderman's $\eta$ in equation~\ref{scaling}.\cite[p. 4]{levi_oz-RMT}}
\end{figure}

Levi and Oz \cite{levi_oz-RMT} study such datasets using ``tools from statistical physics and Random Matrix Theory (RMT)\footnote{Random matrices are matrices whose elements are randomly sampled from a given probability distribution.  RMT focuses primarily  on  behaviors of such matrices as they ``get big'' in analogy to the study of limiting behavior of standard random variables.} to reveal their underlying structure.'' \cite[p. 1]{levi_oz-RMT} They study the eigenvalue spectra  of matrices representing samples from the datasets:
\begin{equation}\label{Gram_matrices}
\Sigma_M = \frac{1}{M}X X^T
\end{equation} 
where $X\in \mathbb{R}^{d \times M}$ where $d$  the dimension of the image vectors, and $M$ is the number of samples. The matrix $\Sigma_M$ is an empirical covariance (Gram) matrix.\footnote{\label{RMTeigen}RMT studies (among other things) the distributions of the eigenvalues of covariance matrices $M$. The distribution of these eigenvalues yields information  about the correlational structure between the matrix elements. In the cases under discussion, this is information about the correlational structure \textit{in the images}. The next section (\ref{weight-matrices}) uses RMT to study the emergence of correlational structure \textit{in the weight matrices for layers of nodes/neurons in DNNs} as they are trained on the datasets examined by Levi and Oz (among others).}  Their empirical investigations show that the 
\begin{quote}spectrum of $\Sigma_M$ for various datasets can be separated into a set of large eigenvalues $(\mathcal{O}(10))$, a bulk of eigenvalues which decay as a power law $\lambda_i\sim i^{-1-\alpha}$  and a large tail of small eigenvalues which terminates at some finite index $n$. \ldots \textit{The bulk of the eigenvalues \ldots can be understood as representing the correlation structure of different features amongst themselves \ldots} . \cite[p. 2, Emphasis added.]{levi_oz-RMT} \end{quote}

 \noindent Furthermore, Levi and Oz \textit{explicitly} refer to Ruderman's scaling exponent in this context.

Their investigations reveal that despite the fact that the real world image datasets are quite varied\footnote{For example, the data include ``natural'' images as well as images of human made \textit{artifacts}. This strengthens Ruderman's claim about power law scaling in images.}, they nevertheless exhibit \textit{universal} power law scaling that is quite distinct from data generated by sampling from an uncorrelated normal distribution---what they call ``uncorrelated Gaussian Data (UGD).''  \cite[p. 2]{levi_oz-RMT}. As evidenced in figure~\ref{scree}, this universality is striking.\footnote{Cattell \cite{scree} introduces the ``scree plot'' as follows:  The bulk of the eigenvalues exhibit power law scaling but the plot tails off:  The  tail  of small (non-bulk) eigenvalues refers to ``[the] straight end portion we began calling the \textit{scree}---from the straight line of rubble and boulders which forms at the pitch of sliding stability at the foot of a mountain. The initial implication was that this scree represents a `rubbish' of small error factors.''} Given that Gaussian distributions are fully characterized by their first two moments (the mean and variance), the deviation from UGD that Levi and Oz find is an indication that statistics beyond the first two moments are needed to classify images within the datasets.

As noted the eigenvalue bulk exhibits power law decay:  $\lambda_i \sim i^{-1-\alpha}$.\footnote{The $\lambda_i$ are the eigenvalues of the correlation or covariance matrix of the dataset. $i$ is the rank of the eigenvalue, $i=1$ being the largest, followed by $i=2$ \ldots. $\lambda_i$ is the variance  of the $i^{th}$ principal component.\label{eigen}} All of the real world datasets have $\alpha \leq 1/2$ where the value of $\alpha$ reflects the strength of the correlations in the covariance matrices. This result, shown in figure~\ref{scree} corresponds precisely to the scaling result Ruderman finds: There is no special $i$ and no special physical frequency $k$; i.e., there is scale invariace in either $i$ or $k$. The scaling is thus universal.

The plots in figure~\ref{scree} are log-log plots where the straight lines of the same slope correspond to the same power law scaling. By comparison, the horizontal line for the UGD  data (uncorrelated meaning no statistical structure) gives an indication of just how (statistically) structured the real world datasets are.
As  noted, this is a striking result that surely must figure in an explanation for how and why the DNNs can become so accomplished at image recognition.\footnote{This, to our minds, is much more compelling evidence than that suggested by  Lin et al.\ \cite{lin-deep-cheap}.} The work discussed in this section demonstrates the presence of higher order statistical structure in real world data. This structure is, of course, available to the DNNs. Whether or not DNNs actually use this structure and, if so, which structural features they employ is a further question to which we now turn. The next section \ref{weight-matrices} surveys  some empirical investigations that help to answer these questions.

\subsection{RMT and the Statistics of Layer Weight Matrices During Training\label{weight-matrices}}

An extensive empirical evaluation of state of the art DNNs provides compelling evidence that  the weight matrices for various layers  of the DNNs undergo  changes in statistics during training.  This work is described in ``Implicit Self-Regularizaiton in Deep Neural Networks: Evidence from Random Matrix Theory and Implications for Learning.'' \cite{Martin_Mahoney_rmt}
 In fact, Martin and Mahoney argue that ``the weight matrices `learn' the correlations in the data.''  \cite[p. 29]{Martin_Mahoney_rmt}

Martin and Mahoney represent eigenvalue spectra of the layer weight matrices\footnote{They call it the energy landscape  ($E_{DNN}$) in loose analogy with RMT analyses of Hamiltonians in dynamical systems. It is a mathematical description of the DNN's layers and is also known as the ``optimization function''---the function that is optimized under training using SGD.} of a ``typical'' DNN having $L$ layers with activation functions $h_l(\cdot)$, weight matrices per layer $\boldsymbol{W}_l$, and biases $\boldsymbol{b}_l$ as:
\[E_{DNN} = h_L(\boldsymbol{W}_L \times h_{L-1}(\boldsymbol{W}_{L-1} \times h_{L-2}( \cdots) + \boldsymbol{b}_{L-1}) + \boldsymbol{b}_L).\]
They study the weight matrices $\boldsymbol{W}_l$ \textit{before, during, and after} training on various datasets for a wide range of actual DNNs. Specifically, they ``analyze the distribution of eigenvalues, i.e., the Empirical Spectral Density (ESD), $\rho_N(\lambda)$, of the correlation matrix $\boldsymbol{X} = \boldsymbol{W}^T\boldsymbol{W}$ associated with the layer weight matrix $\boldsymbol{W}$.'' \cite[p. 5]{Martin_Mahoney_rmt} These are, again,  empirical covariance  matrices, though in this case for the \textit{layer weights}, and \textit{not} for the samples from the datasets  studied by Levi and Oz as discussed in section~\ref{scale_data}.

Given a  dataset $\mathcal D$ of labeled data $\{d_i,y_i\} \in \mathcal D$ the goal of machine learning is to minimize the loss\footnote{The loss $\mathcal L$ is the cross-entropy, which measures the mismatch between the predicted probability distribution of the DNN outputs and the ground-truth distribution of actual image labels.}, $\mathcal L$, between the $E_{DNN}$ and the labels $y_i$:
\begin{equation}
 \min_{W_l,b_l}\mathcal{L}\left(\sum_i E_{DNN}(d_i) - y_i\right).
\end{equation}
Typically, to avoid overfitting, this requires regularization by \textit{explicitly} adding a term that selects for small norm functions. That is, terms are added that ``shrink the norm(s) of the $\boldsymbol{W_l}$ matrices'' \cite[p. 5]{Martin_Mahoney_rmt}
as follows: 
\begin{equation}\label{norm}
\min_{W_l,b_l}\mathcal{L}\left(\sum_i E_{DNN}(d_i) - y_i\right) + \alpha \sum_l \| \boldsymbol{W_l} \|. 
\end{equation}
Martin and Mahoney show that large DNNs trained on the image datasets, effectively implement an \textit{implicit self-regularization} that they call ``Heavy-Tailed Self-Regularization.'' \cite[p. 6]{Martin_Mahoney_rmt}. They demonstrate that the explicit introduction  of a regularizing norm (the second term in equation~\ref{norm}) is not required for large, state of the art, DNNs to generalize.  We think that it is in this sense that Martin and Mahoney understand the claim that the weight matrices  are \textit {implicitly} learning ``the correlations in the data.'' That is to say, the weight matrices of the DNNs under training take on the same heavy tail statistics that are present in the datasets as we now discuss.

\begin{figure}[h]\centering
\includegraphics[width=.99\textwidth,trim=0in 0in 0in 0in,clip]{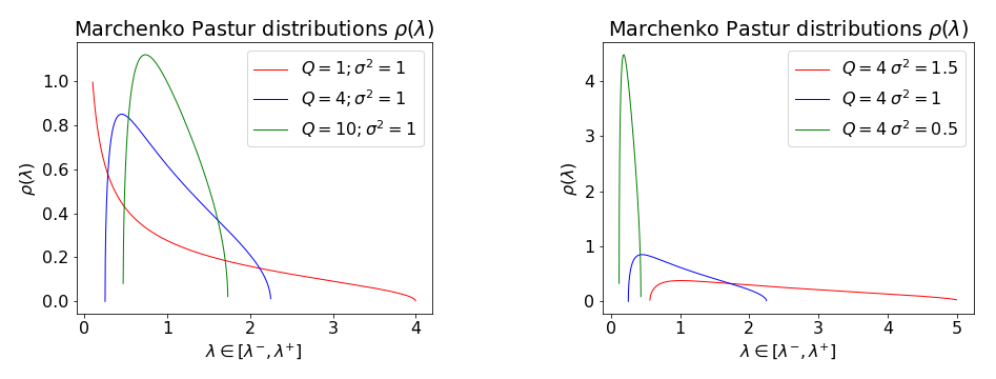}
\caption{\label{MP}Mar\v{c}henko-Pastur (MP) Distributions  \cite[p.14]{Martin_Mahoney_rmt}}
\end{figure}

Random Matrix Theory provides Law of Large Numbers-like and Central Limit Theorem-like results for matrices.  It yields unique results for both square and rectangular matrices. As Martin and Mahoney note, in DNNs square weight matrices are rare. Typically, the number of parameters ($N$) is greater than the number of samples ($M$).  Much work in RMT has focused on a class of matrices that are members of a so-called \textit{Universality class of Gaussian Distributions}: Given a matrix $\boldsymbol{W}$   assume that the elements $W_{i,j}$ are drawn from a Gaussian distribution:
\[W_{i,j}\sim N(O,\sigma^2_{mp}).\]  
Under these assumptions, RMT shows that the  Empirical Spectral Density\footnote{This is the distribution of the eigenvalues, $\lambda$, of the covariance matrix. See footnote~\ref{eigen} above.} (ESD) of the correlation matrix $\boldsymbol{X} = \boldsymbol{W}^T\boldsymbol{W}$: \[\rho_N(\lambda) := \frac{1}{N}\sum_{i=1}^M \delta (\lambda - \lambda_i),\]   has the Mar\v{c}henko-Pastur (MP) distribution as its limiting form  (as $N\to\infty$, with aspect ratio $Q=N/M  \geq 1$ fixed\footnote{This is the ratio of the number of rows ($N$)---the depth of the layer weight matrix---to the number of columns ($M$)---its width.})  \cite[p. 14]{Martin_Mahoney_rmt}:

\begin{equation}
 \lim_{N\to \infty} \rho_N(\lambda) = 
\begin{cases}
\frac{Q}{2\pi\sigma_{mp}^2} \frac{\sqrt{(\lambda^+ - \lambda)(\lambda - \lambda^-)}}{\lambda}, & \text{if $\lambda \in [\lambda^-,\lambda^+]$}\\
0, & \text{otherwise.}
\end{cases}
\end{equation}
These distributions are shown in figure~\ref{MP}.  In effect, these are the RMT analogs (for matrices) of various (uncorrelated) Gaussian/normal distributions in ordinary probability theory. Deviations from these distributions measure the extent to which matrices have additional structure.

\begin{figure}[h]\centering
\includegraphics[height=.46\textheight,trim=0in 0in 0in 0in,clip]{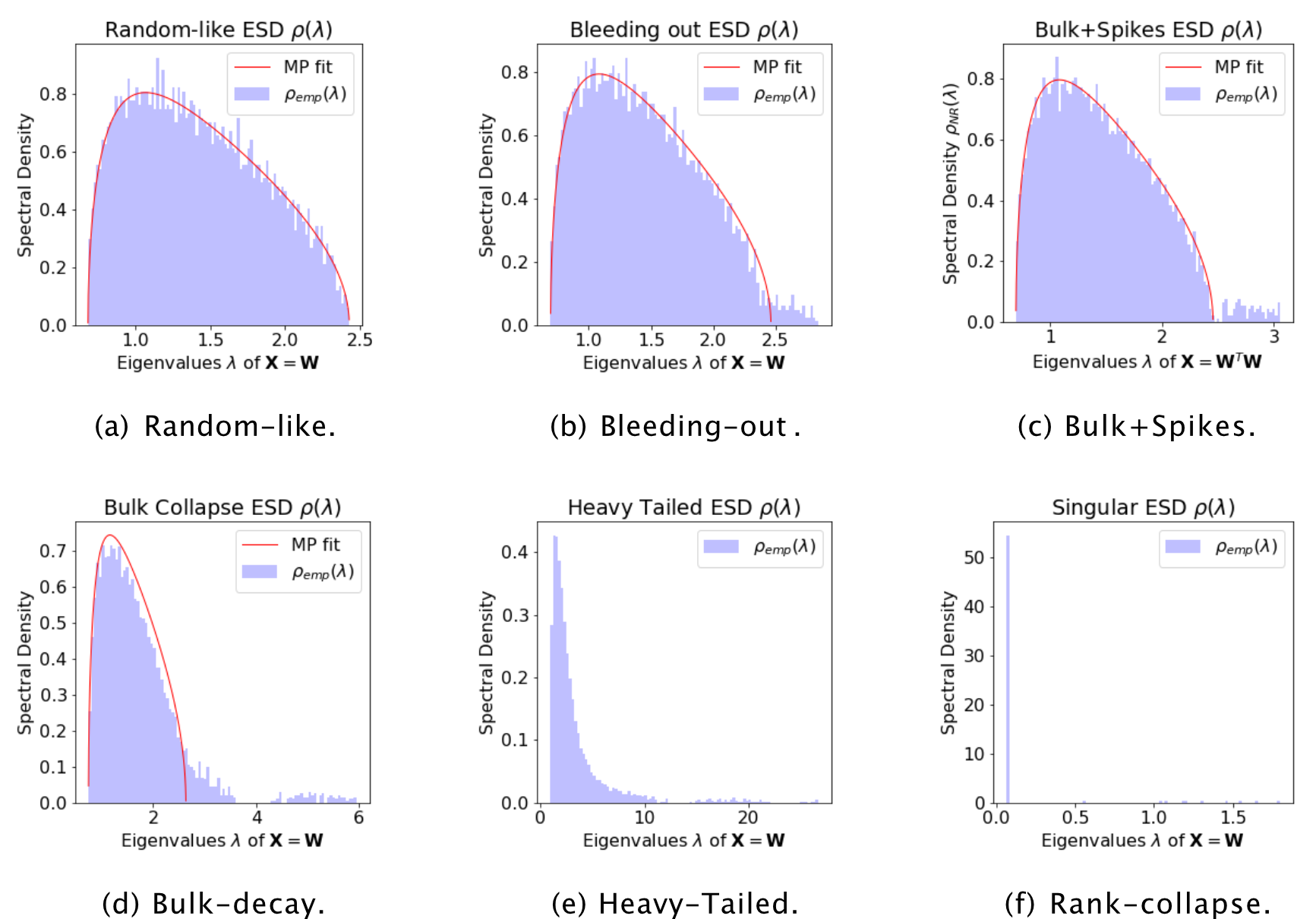}
\caption{\label{to_fat}Taxonomy of Trained Models. Changing RMT statistics for Weight Matrix Spectral Densities \cite[p.32]{Martin_Mahoney_rmt}}
\end{figure}

In their investigation of the statistics of the weight matrices for fully connected layers in \textit{trained} state-of-the-art DNNs, Martin and Mahoney find ``profound deviations from traditional [MP or Gaussian based] RMT.''  And they find that these DNNs ``are reminiscent of strongly-correlated disordered systems that exhibit Heavy-Tailed behavior.'' \cite[p. 29]{Martin_Mahoney_rmt}  This  is illustrated in figure~\ref{to_fat}. The figure  shows the evolution of the statistics of the ESDs for layer weight matrix $W_l$. Under training using, Stochastic Gradient Descent (SGD), the ESDs evolve from \textit{random-like distributions} (\textit{associated with random weight initialization at the start of training}) with good MP  fit, to Heavy-Tailed distributions\footnote{These distributions correspond to power law distributions of the type seen in the datasets.  See figure~\ref{scree}.}  that correspond to strong correlations in $W_l$ for layers $l$ \textit{at the end of training}.  

The important take-away from this exploration is that \textit{randomly initialized} DNNs begin with Empirical Spectral Density matrices that  have the form of the Mar\v{c}henko-Pastur (MP) distributions as seen in Figure~\ref{to_fat}a, and then, \textit{during training on real world datasets using stochastic gradient descent}, they change/evolve to Heavy-Tailed distributions as shown in Figure~\ref{to_fat}e. Furthermore, this result is robust: It holds for different DNNs and different datasets.

\subsection{Recap\label{recap}}
The recognition that the various training datasets are all members of a universality class exhibiting identical power law scaling suggests that the successes of DNNs at certain tasks may depend, at least in part, on their abilities to discover and utilize specific correlational structures present in \textit{real world} data. 
The empirical investigations by Martin and Mahoney into actual DNNs being trained on such datasets, provides additional evidence that DNNs are exploiting this structure as they are trained using SGD. Moreover, the analyses of Levi and Oz and Martin and Mahoney both agree on form (the exponent) of the power law distributions.

  Prior to constructing and exploring the empirical covariance matrices, equation~(\ref{Gram_matrices}), Levi and Oz  ``pre-process'' the datasets to center  and remove the ``uninformative mean contribution.'' \cite[p. 4]{levi_oz-RMT} In other words, the real world datasets  (MNIST, FMNIST, CIFAR \ldots) as well as the constructed datasets of \textit{correlated} (CGD) and \textit{uncorrelated} Gaussian data (UGD) displayed in figure~\ref{scree}, have been adjusted so as to share the same mean (set to zero).  
  As a result, Levi's and Oz's  investigations into  datasets correlations focus on the second moment of the distributions in the various datasets.  
  They focus on the covariance which captures 2-point correlations as we saw in Section~\ref{obj_scale}.  And, again, figure~\ref{scree} shows the covariance statistic is universal across the real world and CGD data.

Martin's and Mahoney's investigations   show the \textit{evolution} of the Empirical Spectral Densities (ESDs) of correlation matrices for fully connected  layers in actual DNNs from random/Gaussian initialization.  This evolution is the result of neuron weight updating  under training using SGD. 
The results they report show that the ESDs evolve to \textit{take on} the non-Gaussian second moment correlations responsible for the power law statistics \textit{present in the datasets themselves}. 


The investigations examined in this section support  the ideas that (i) the  datasets upon which DNNs are trained exhibit universal non-Gaussian statistics captured in  the second moments of their distributions\footnote{We mean by this  that there are statistical dependencies in such non-Gaussian distributions that are reflected in  second order correlation functions.  Such distributions will also be characterized by  non-zero higher order  correlation functions that  will not follow just from the second order correlations: The second order correlations will not capture the full statistical structure.}, and that (ii)  randomly initialized DNNs will evolve, in training using SGD, to take on the non-Gaussian correlations present in those distributions.  This suggests that perhaps DNNs trained in this manner  will learn \textit{first}  the means and variances of the dataset distributions.\footnote{Some evidence for this is discussed in section~\ref{beyond_2pt}.} But, means and variances ($1$-point and $2$-point correlations) are not themselves sufficient for determining membership in distinct classes within the datasets. In turn, this suggests that DNNs must \textit{subsequently} learn higher order $N$-point correlations in order to classify successfully an image as a member of one class or another.

In section~\ref{cf_method} we described, briefly, a condensed matter/materials science method for upscaling from mesoscale correlational structures to continuum scale material parameters. There we demonstrated the importance of finding higher order correlations for upscaling to approximate continuum values for those parameters.

In the next section we provide  some evidence that DNNs are  \textit{implementing} some version of this correlation function  methodology.  That is, DNNs engaged in image classification (and, we believe, in other tasks as well), are finding higher order correlations---correlations \textit{beyond} the first two moments.
  We motivate this first by examining 2-point and 3-point correlations in the MNIST dataset in section~\ref{npcf}. Following this, in section~\ref{beyond_2pt} we discuss the  work in \cite{Refinetti_Learning_distrib} that further supports this conjecture.

\section{Implementing the Correlation Function\\ Methodology\label{implement}}

This section reports on some  investigations into the following  the two questions: 
\begin{enumerate}[label=Q\arabic*]
\item \label{q1}Can $N$-point correlation functions for ($N > 2$) distinguish members of one class ($x_i\in{\bf x}$) of labelled images from those of another such class in the same dataset\footnote{In the context of  MNIST, the dataset discussed here, the classes are $x_i \in {\bf x} = \{0,1,2,3,4,5,6,7,8,9\}$.}?  
\item Given that higher order correlations are sufficient for distinguishing labelled classes in a dataset, how might the DNNs \textit{actually} go about determining or finding those correlation functions?
\end{enumerate}

\subsection{$N$-point Correlation Functions\label{npcf}}

\begin{figure*}
\begin{multicols}{5}
    \includegraphics[width=\linewidth]{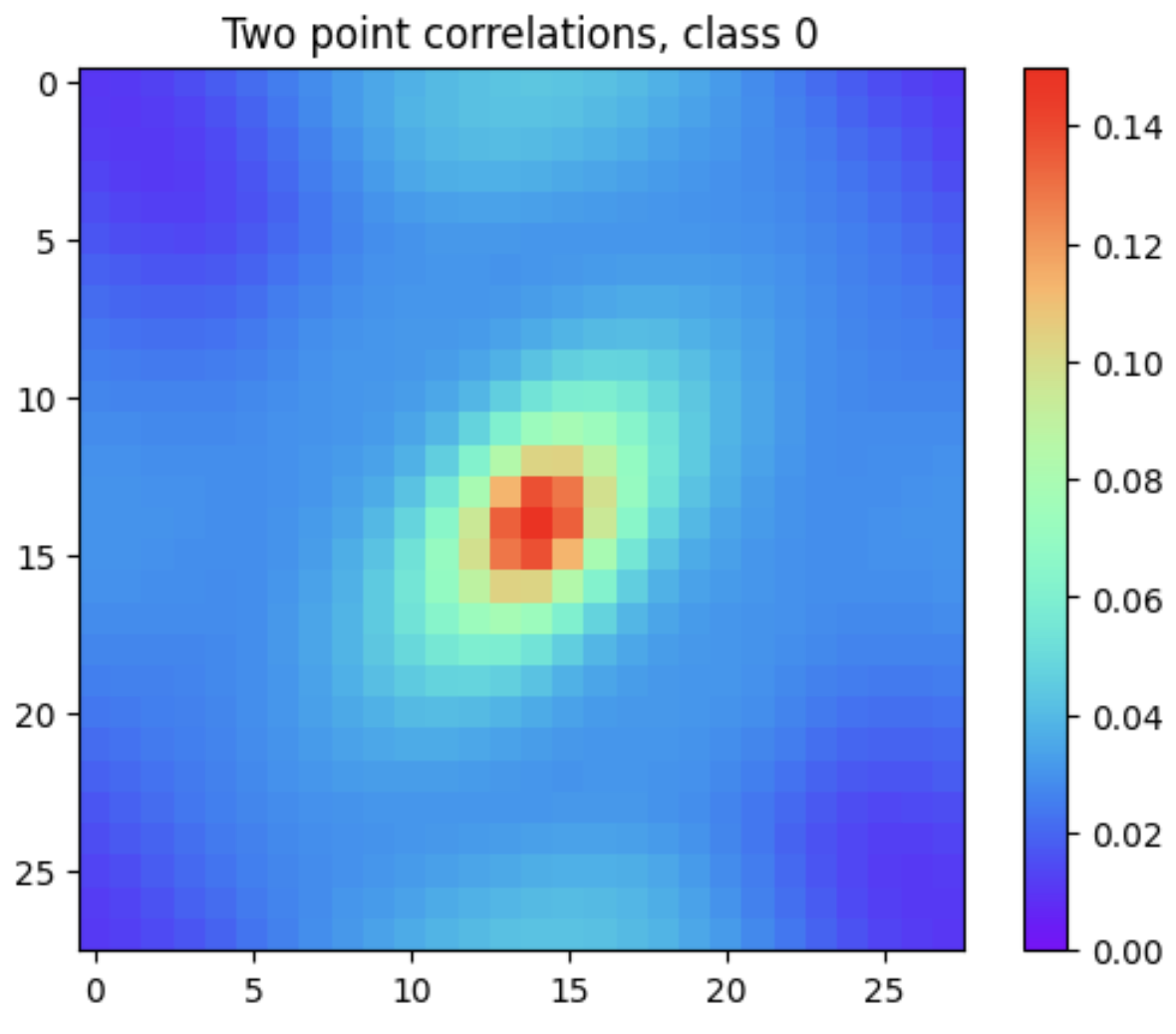}\par 
    \includegraphics[width=\linewidth]{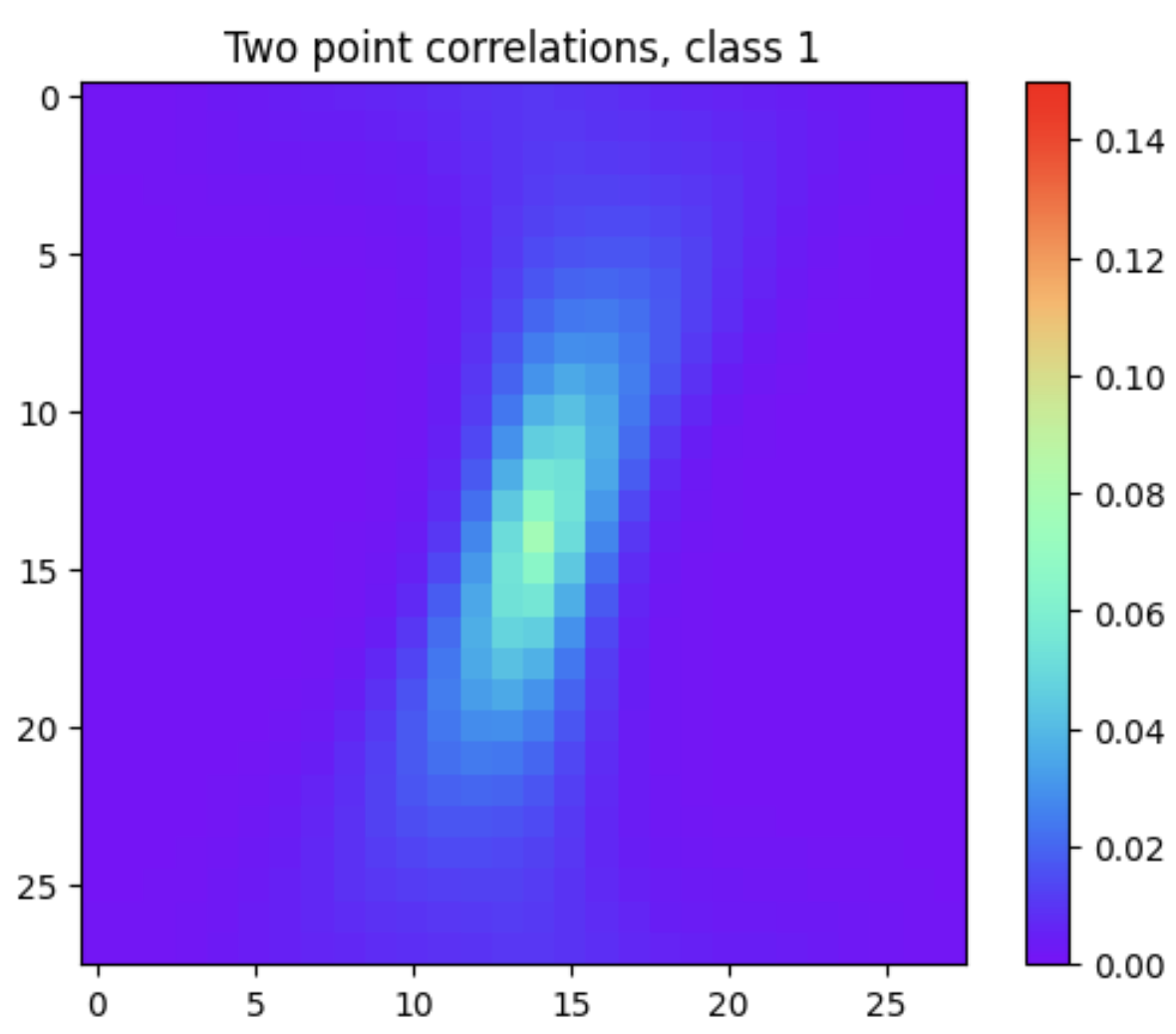}\par 
     \includegraphics[width=\linewidth]{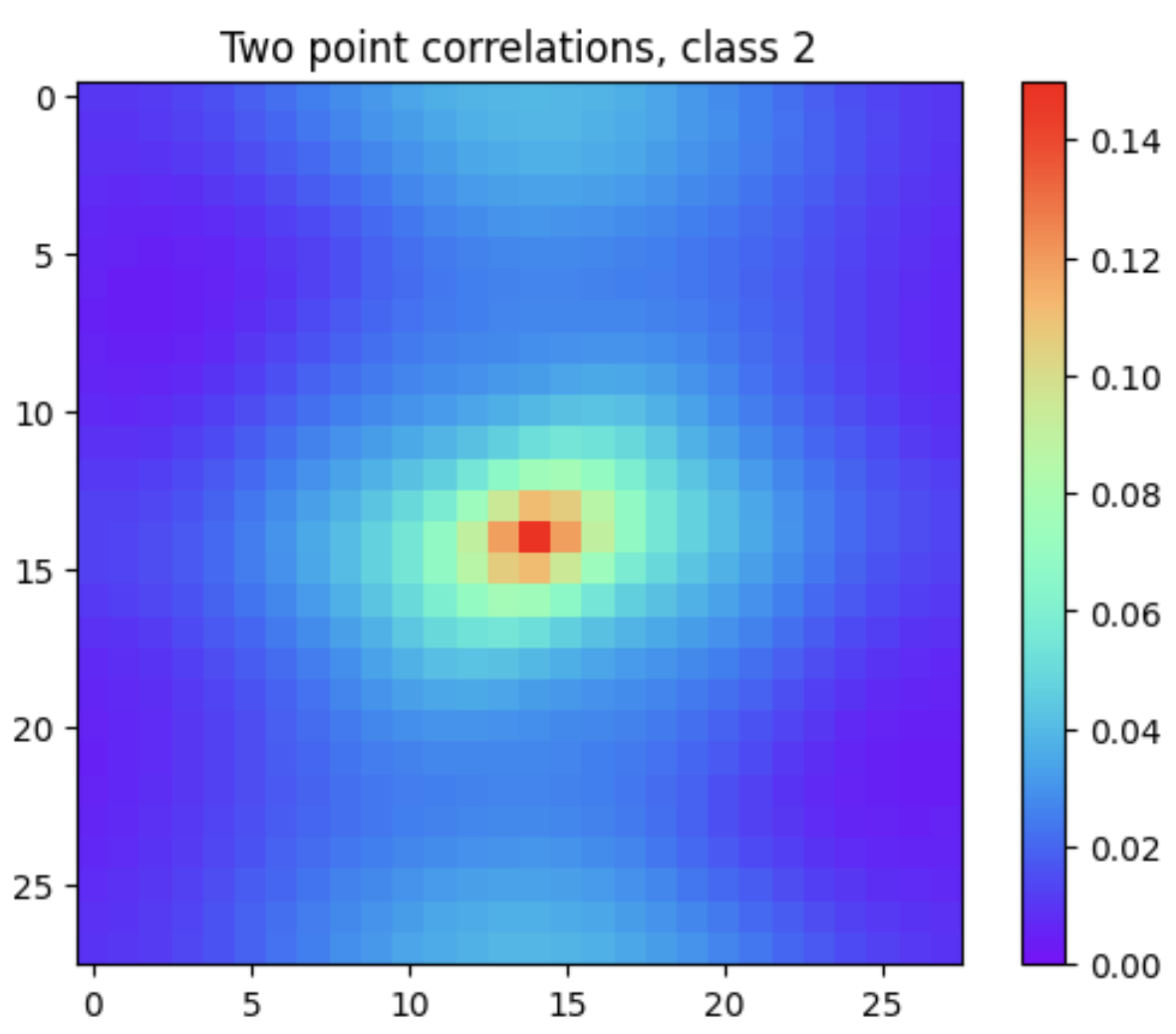}\par
      \includegraphics[width=\linewidth]{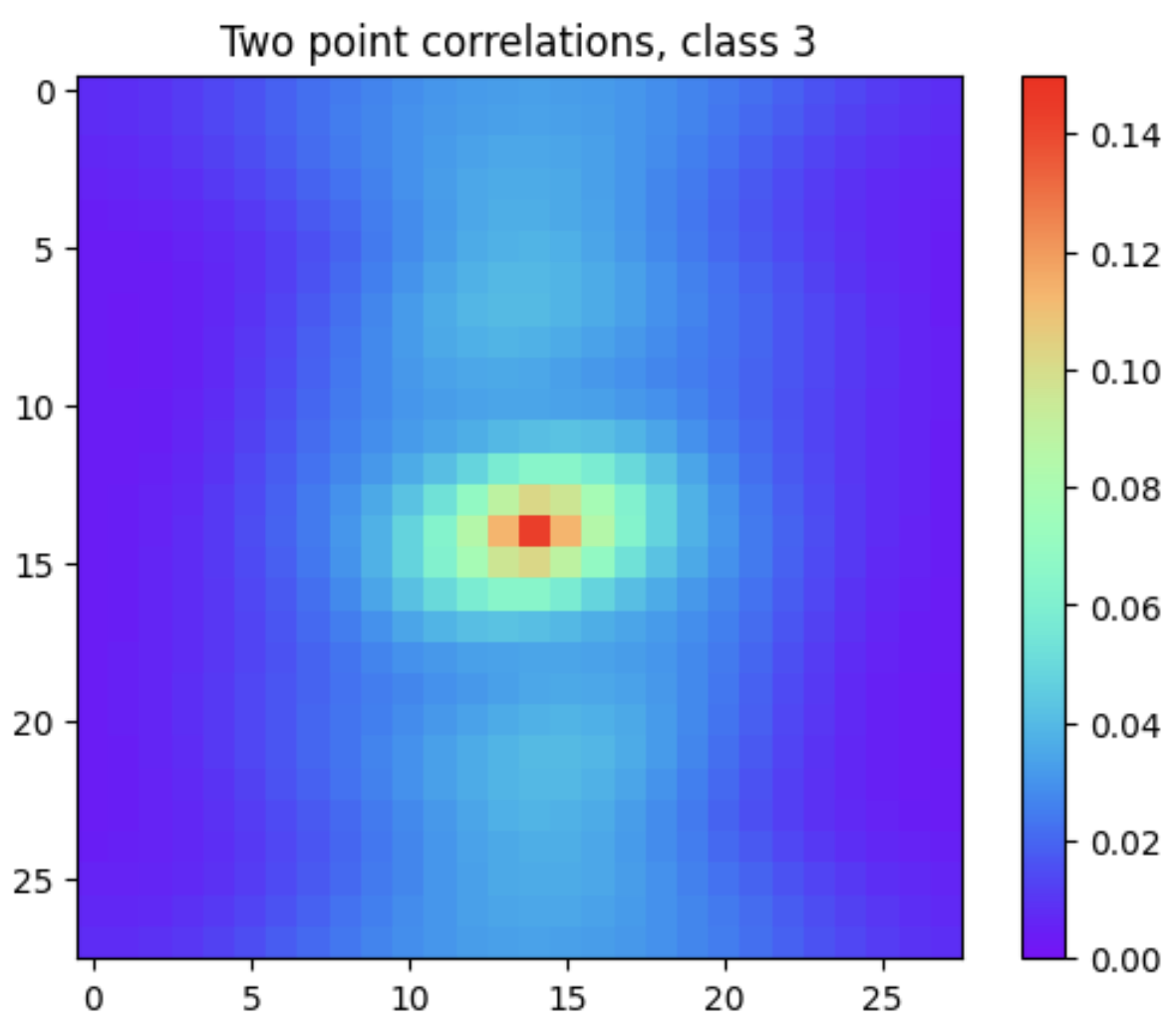}\par
     \includegraphics[width=\linewidth]{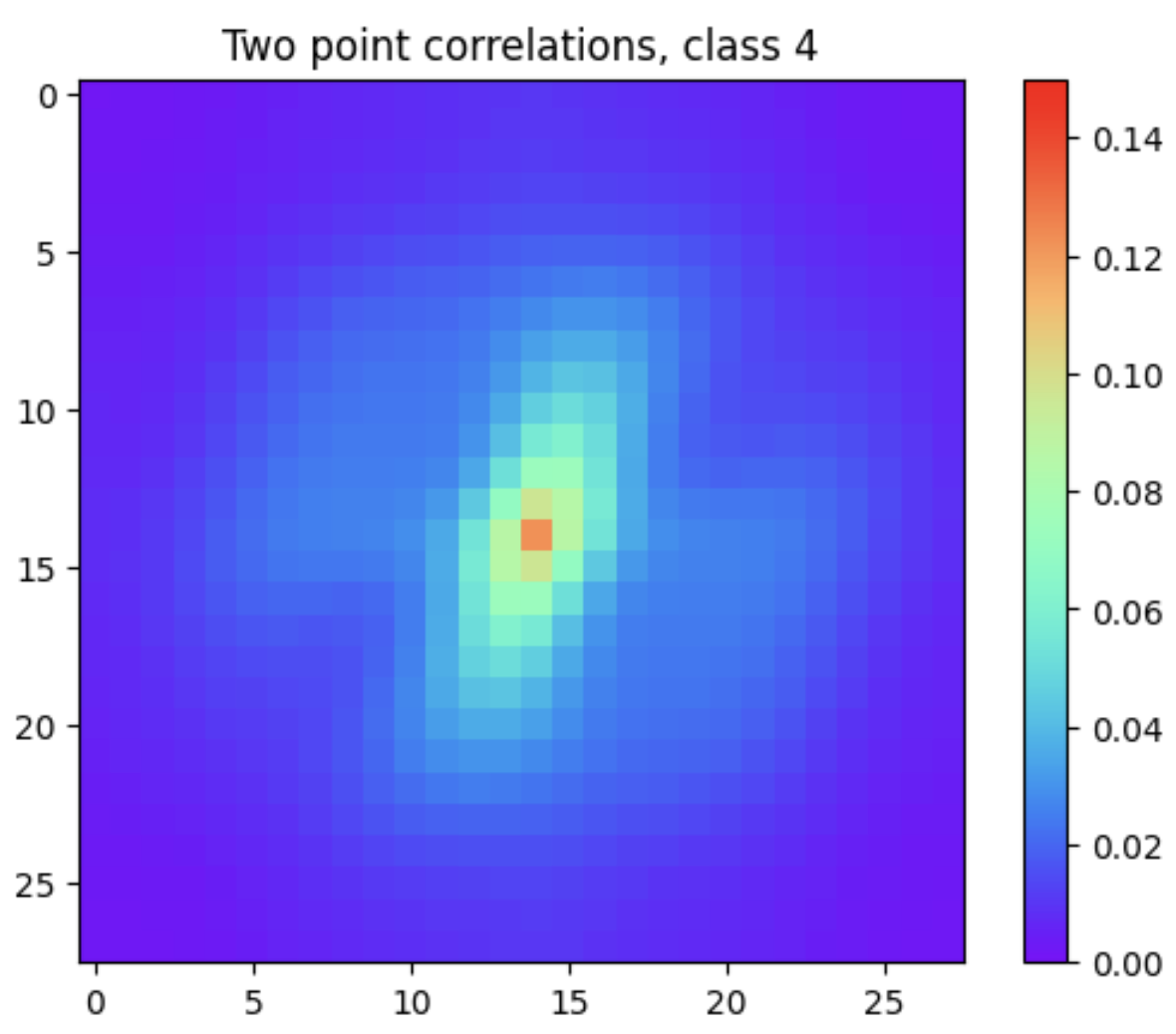}\par
    \end{multicols}
\begin{multicols}{5}
        \includegraphics[width=\linewidth]{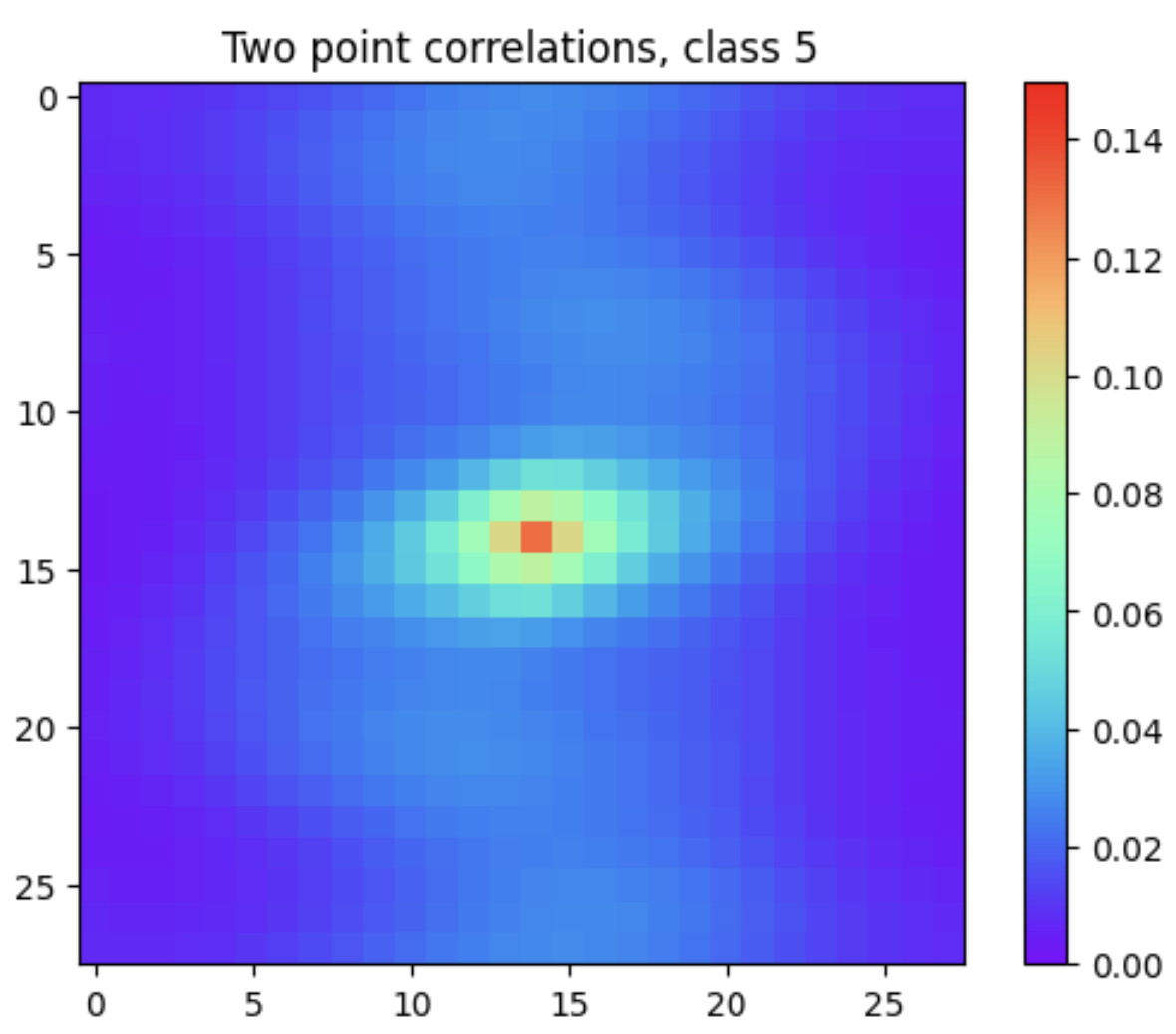}\par
    \includegraphics[width=\linewidth]{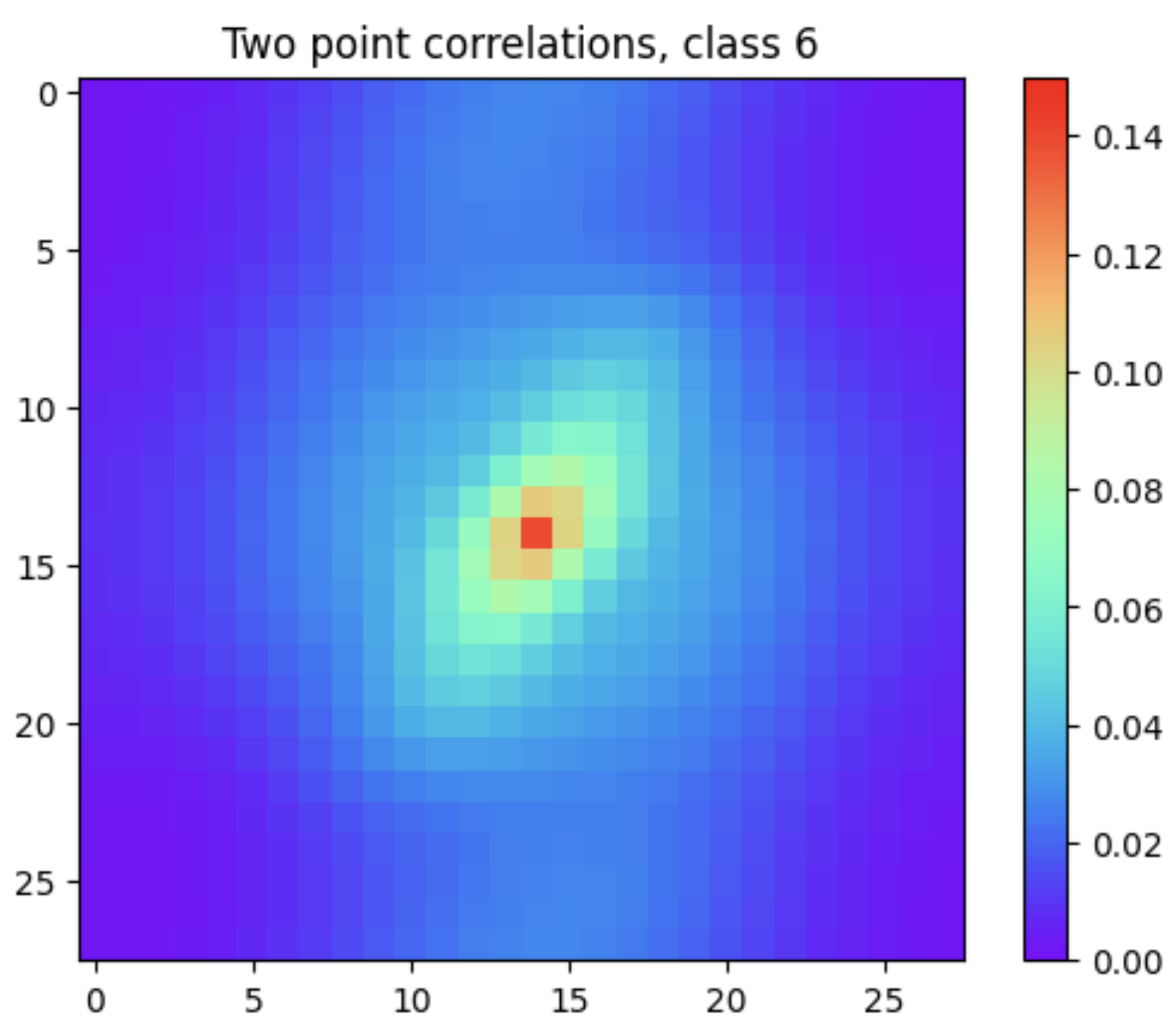}\par
       \includegraphics[width=\linewidth]{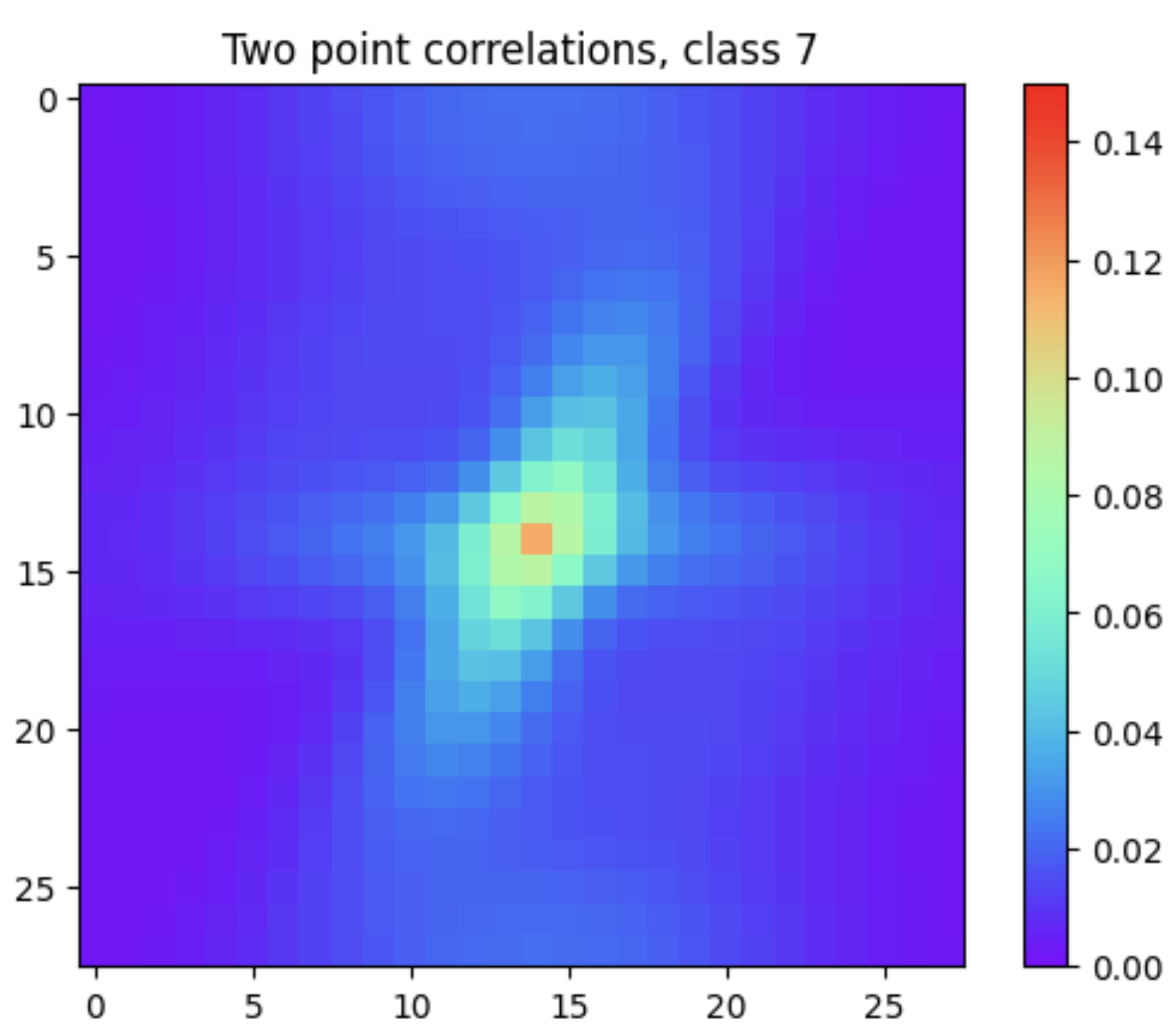}\par
       \includegraphics[width=\linewidth]{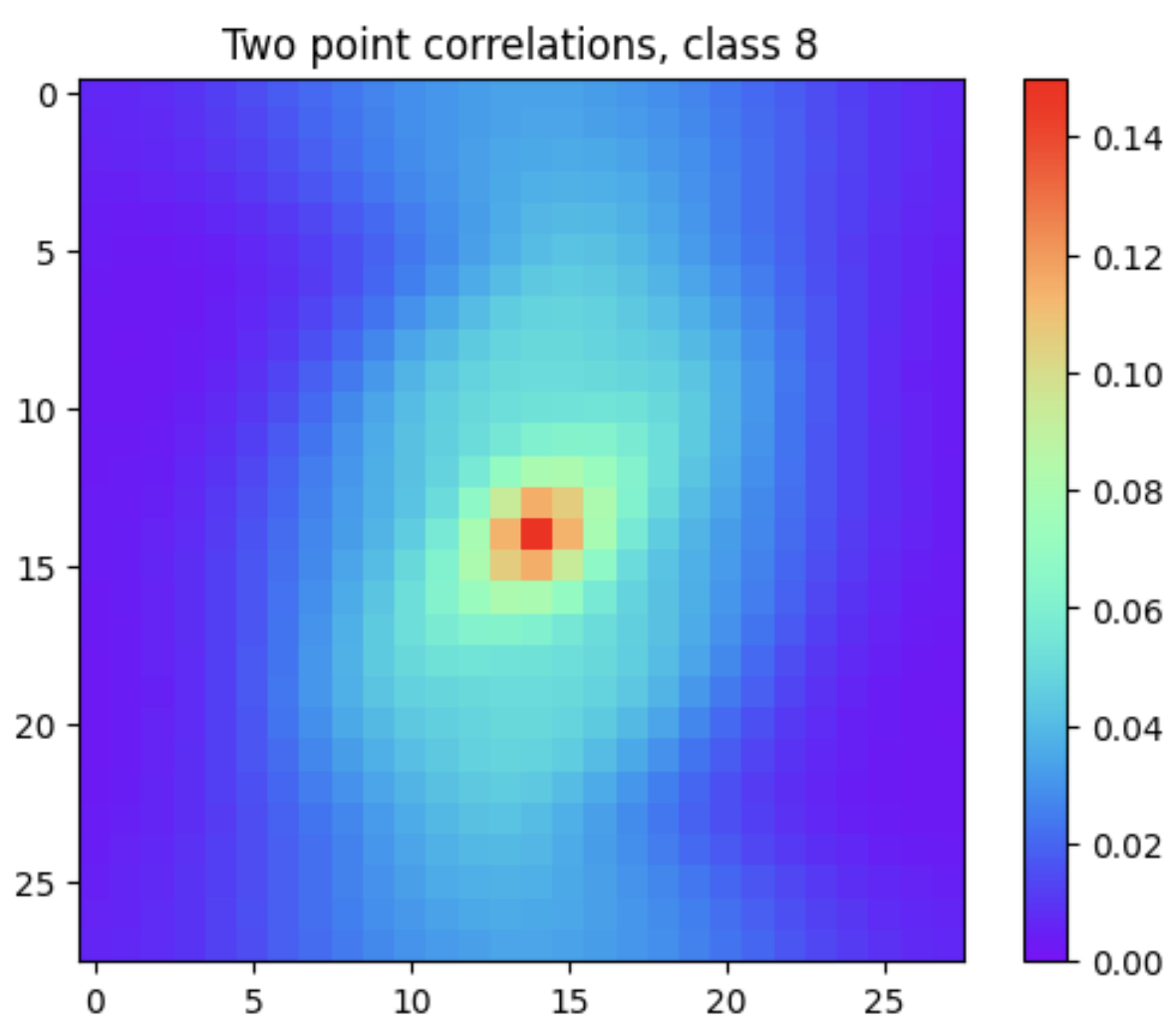}\par
          \includegraphics[width=\linewidth]{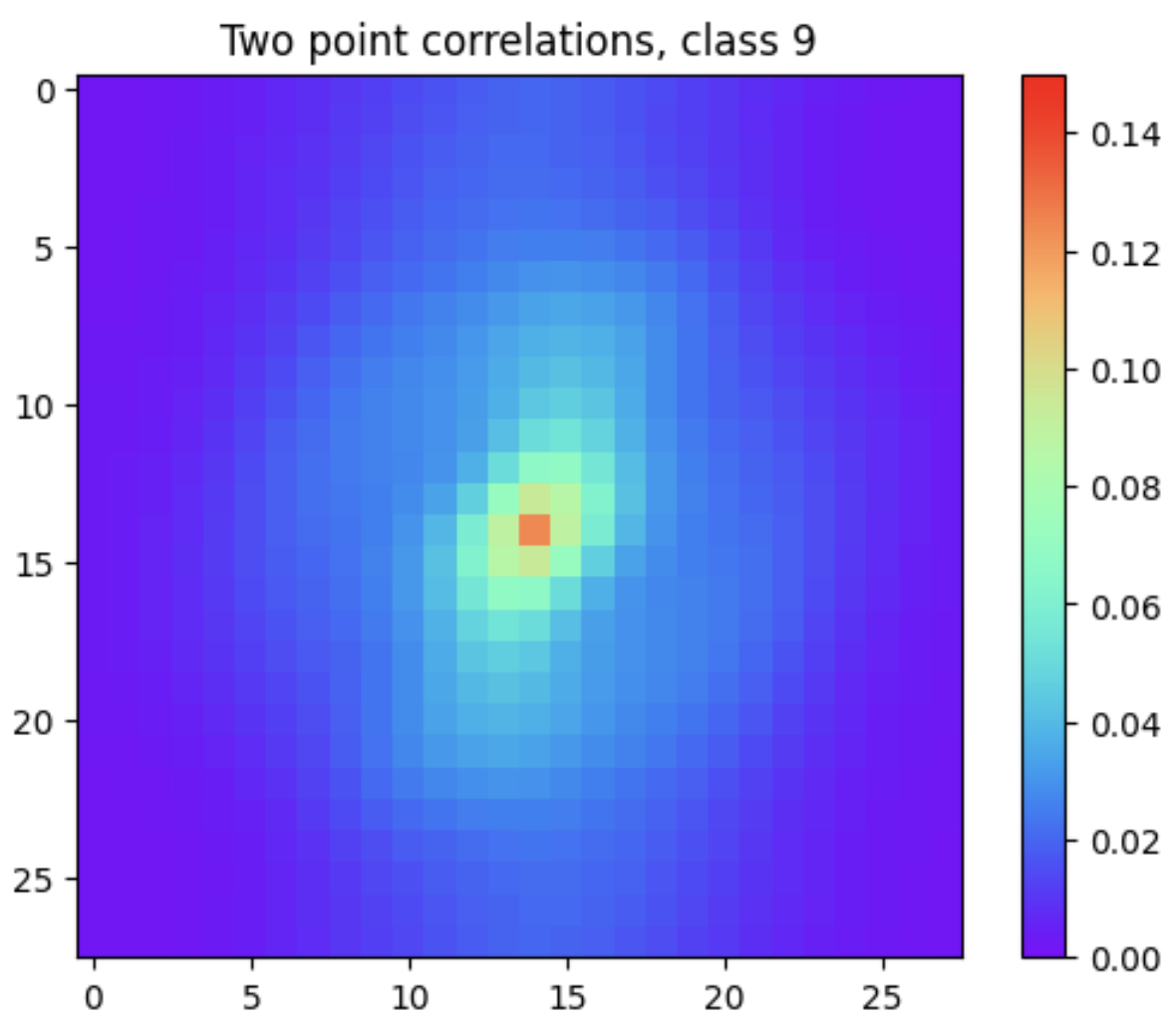}\par
    \end{multicols}

\caption{2-Point Probability Plots.  Scales on Right Shade from Lower Probabilities (Bottom) to Higher Probabilities (Top).}
\label{2-point-plots}
\end{figure*}

\begin{figure*}
\begin{multicols}{5}
    \includegraphics[width=\linewidth]{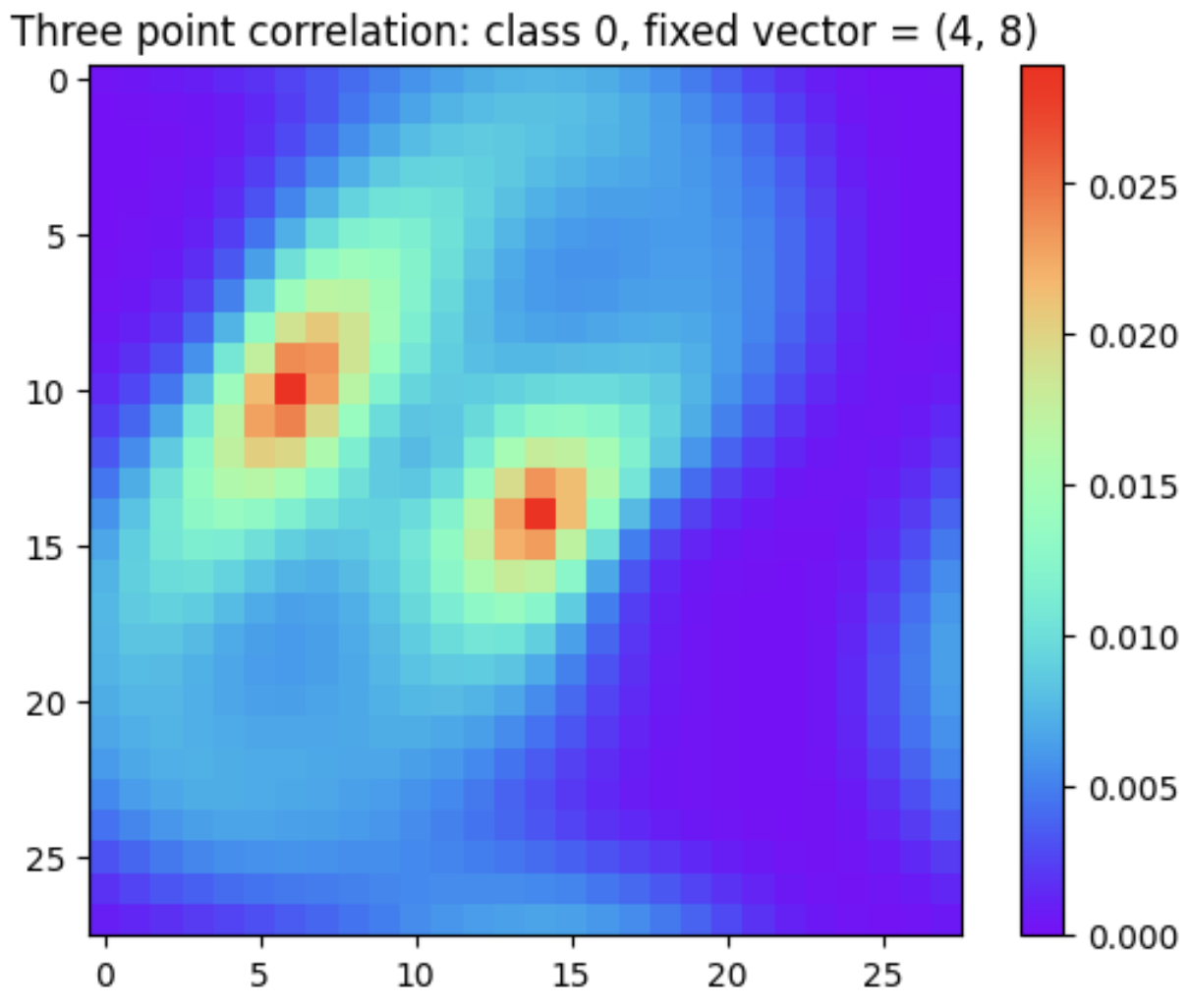}\par 
    \includegraphics[width=\linewidth]{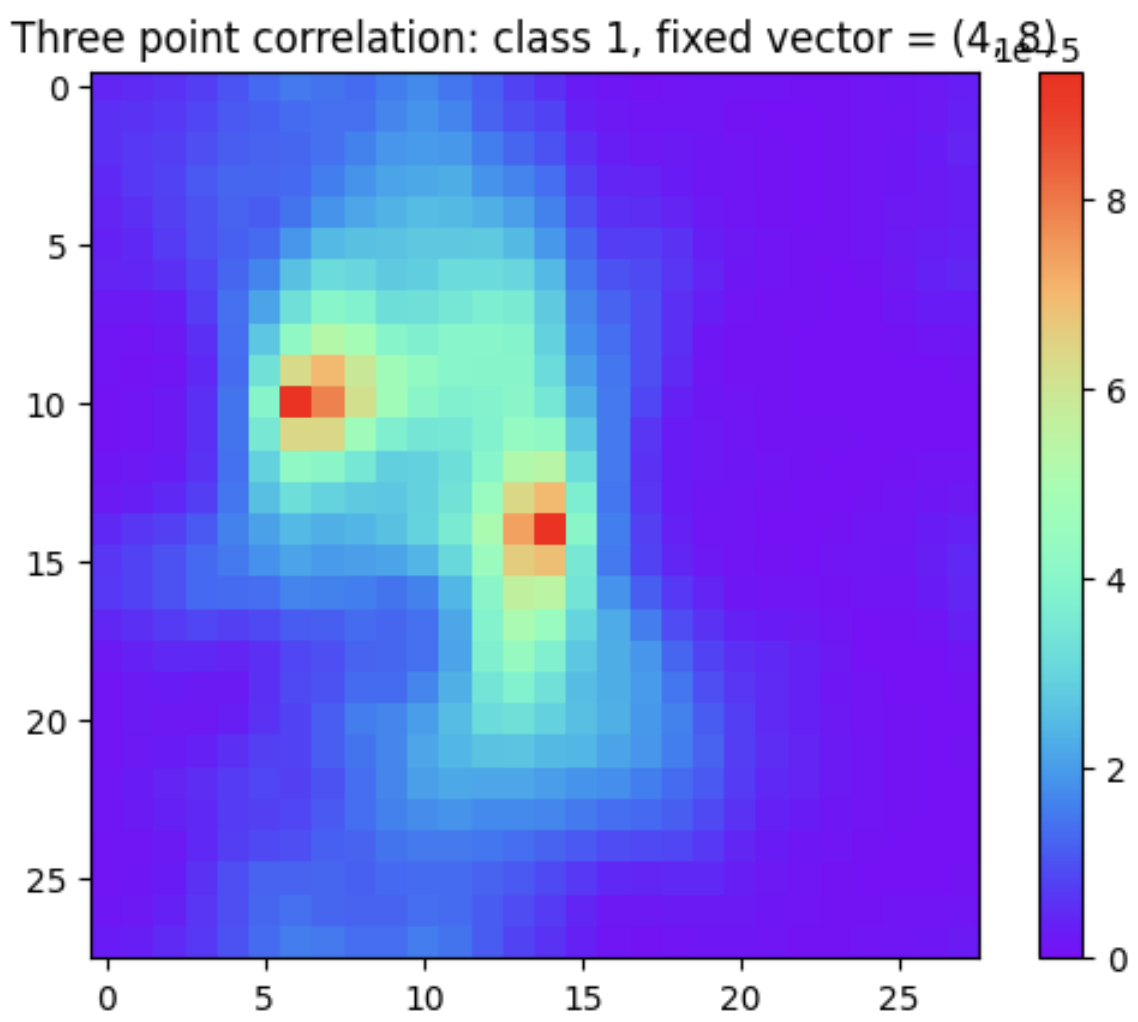}\par 
     \includegraphics[width=\linewidth]{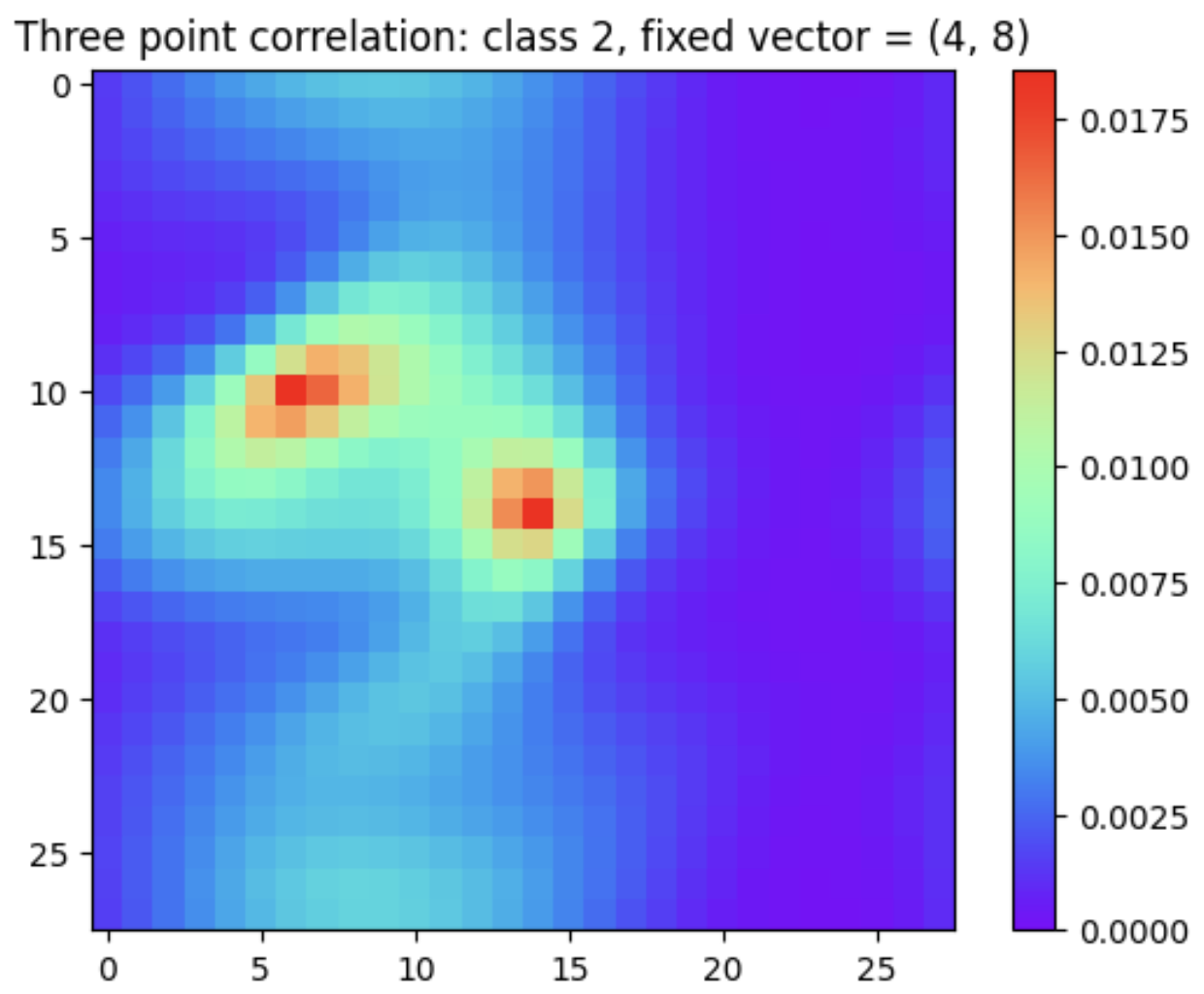}\par
      \includegraphics[width=\linewidth]{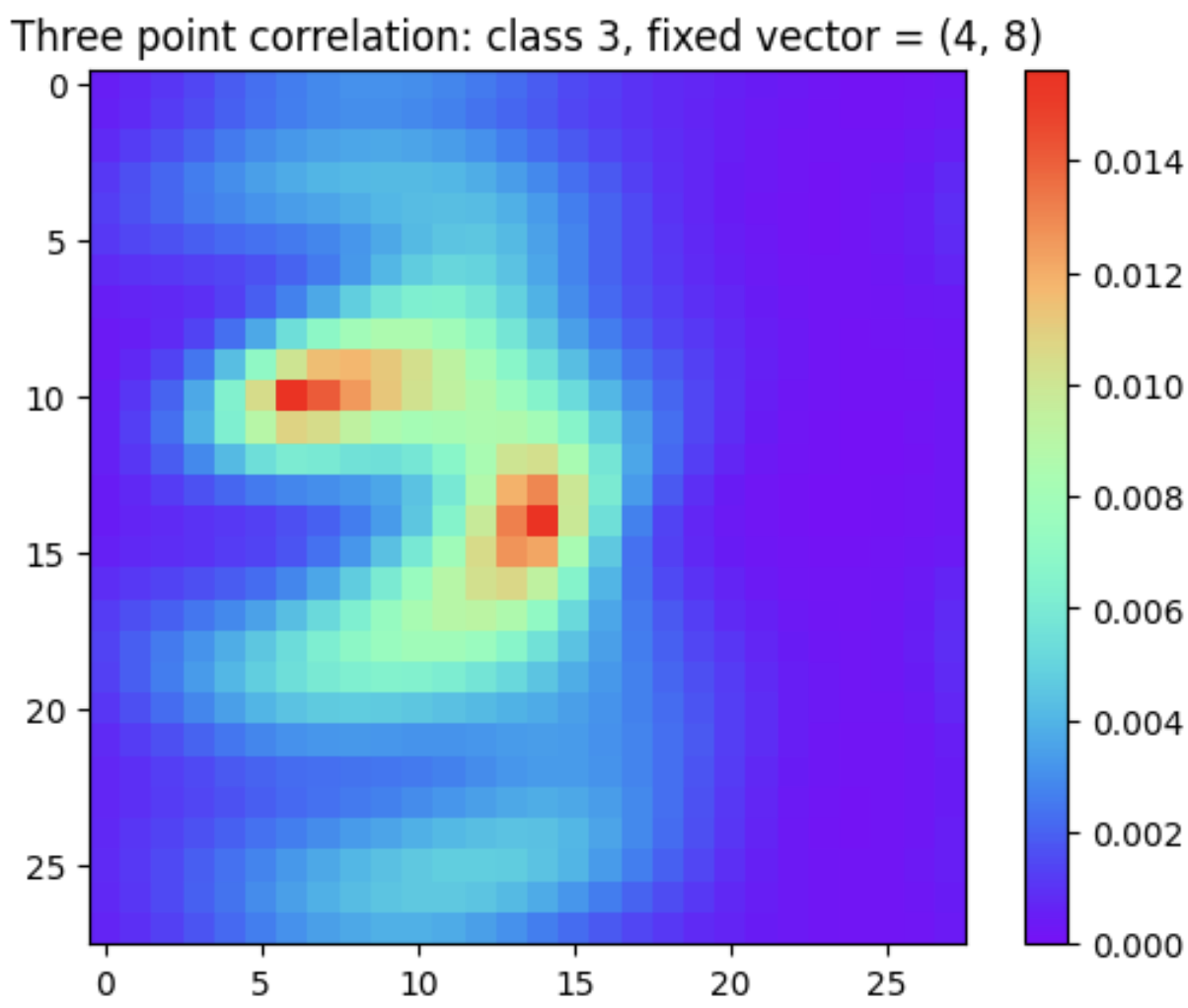}\par
     \includegraphics[width=\linewidth]{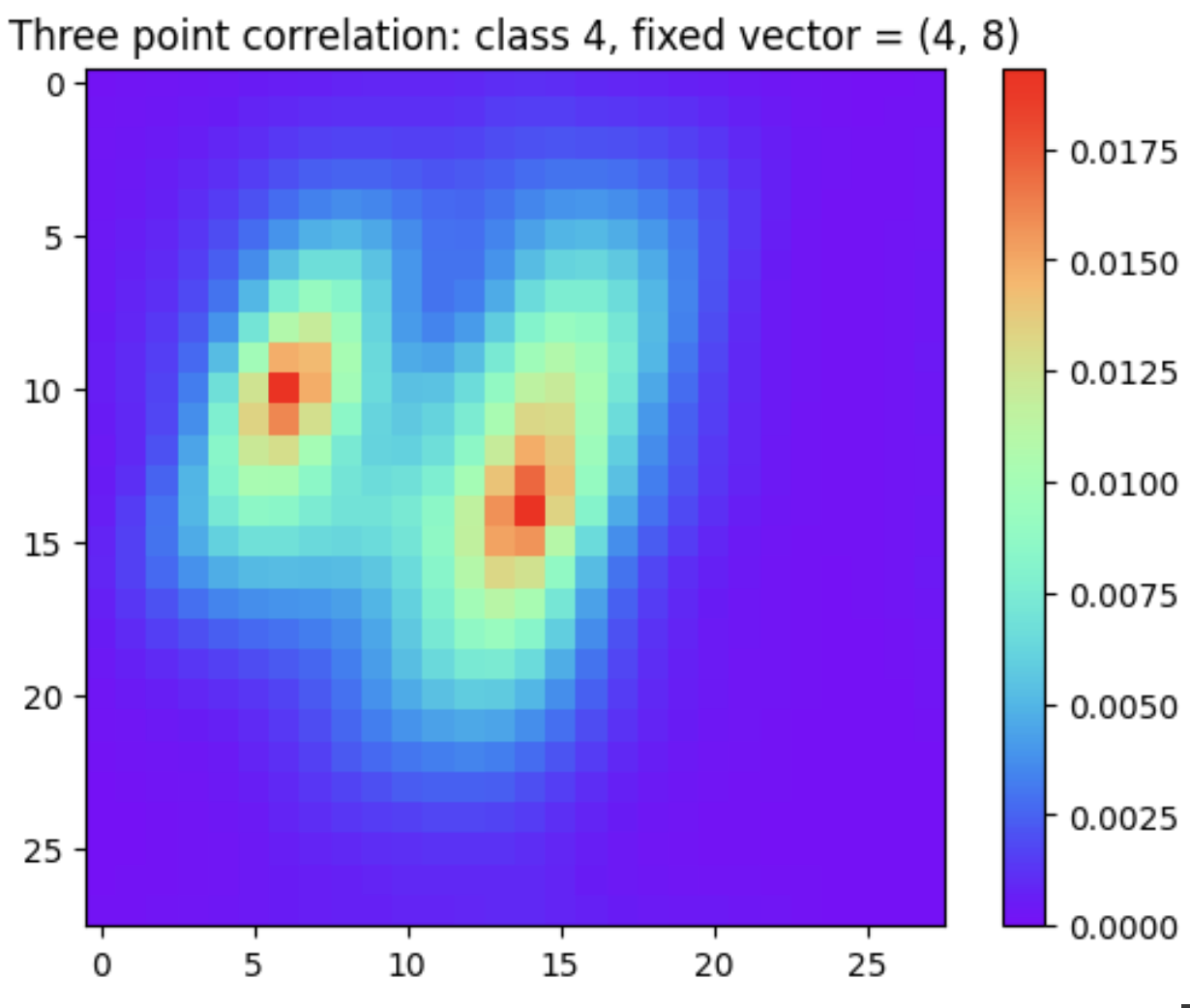}\par
    \end{multicols}
\begin{multicols}{5}
        \includegraphics[width=\linewidth]{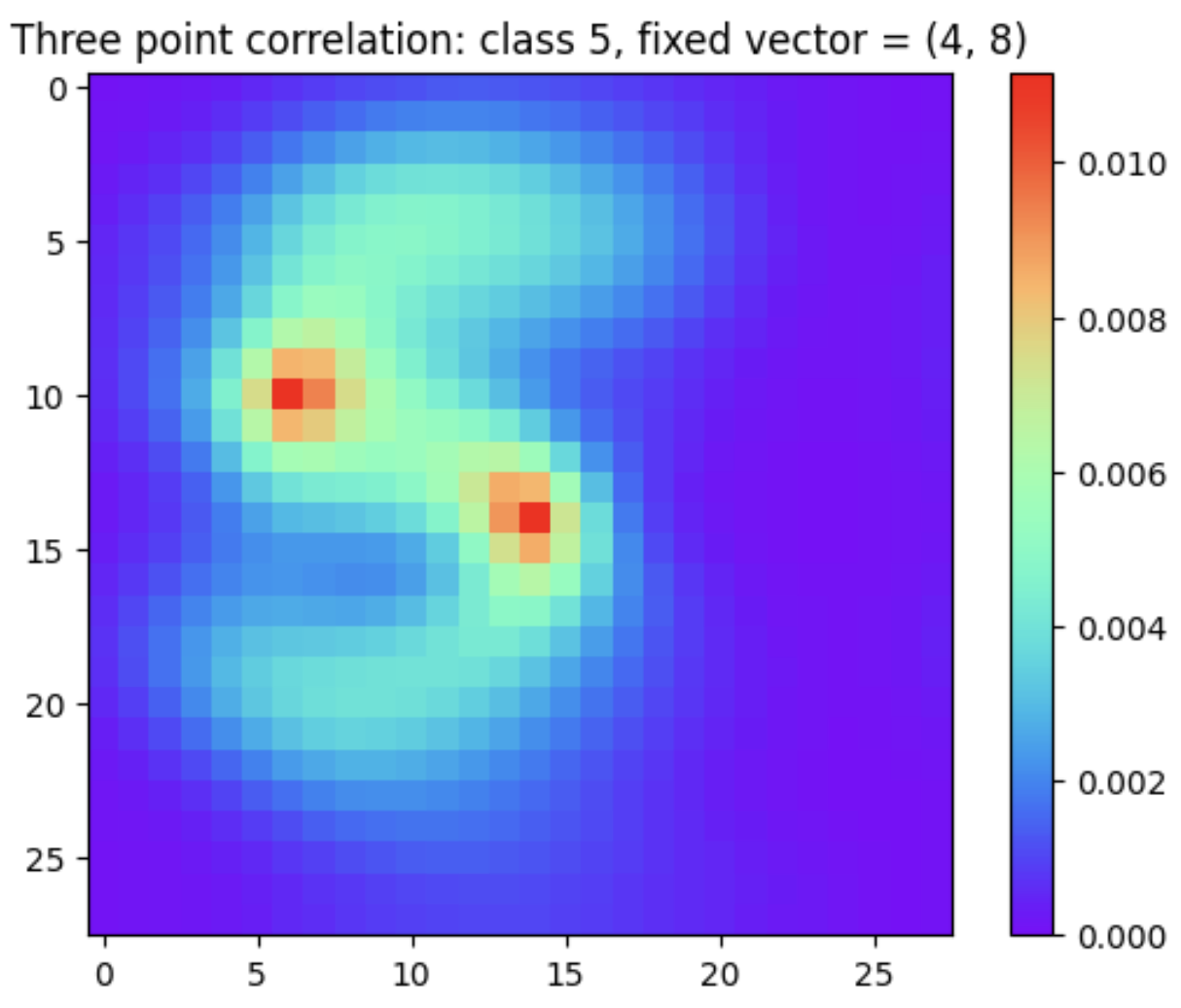}\par
    \includegraphics[width=\linewidth]{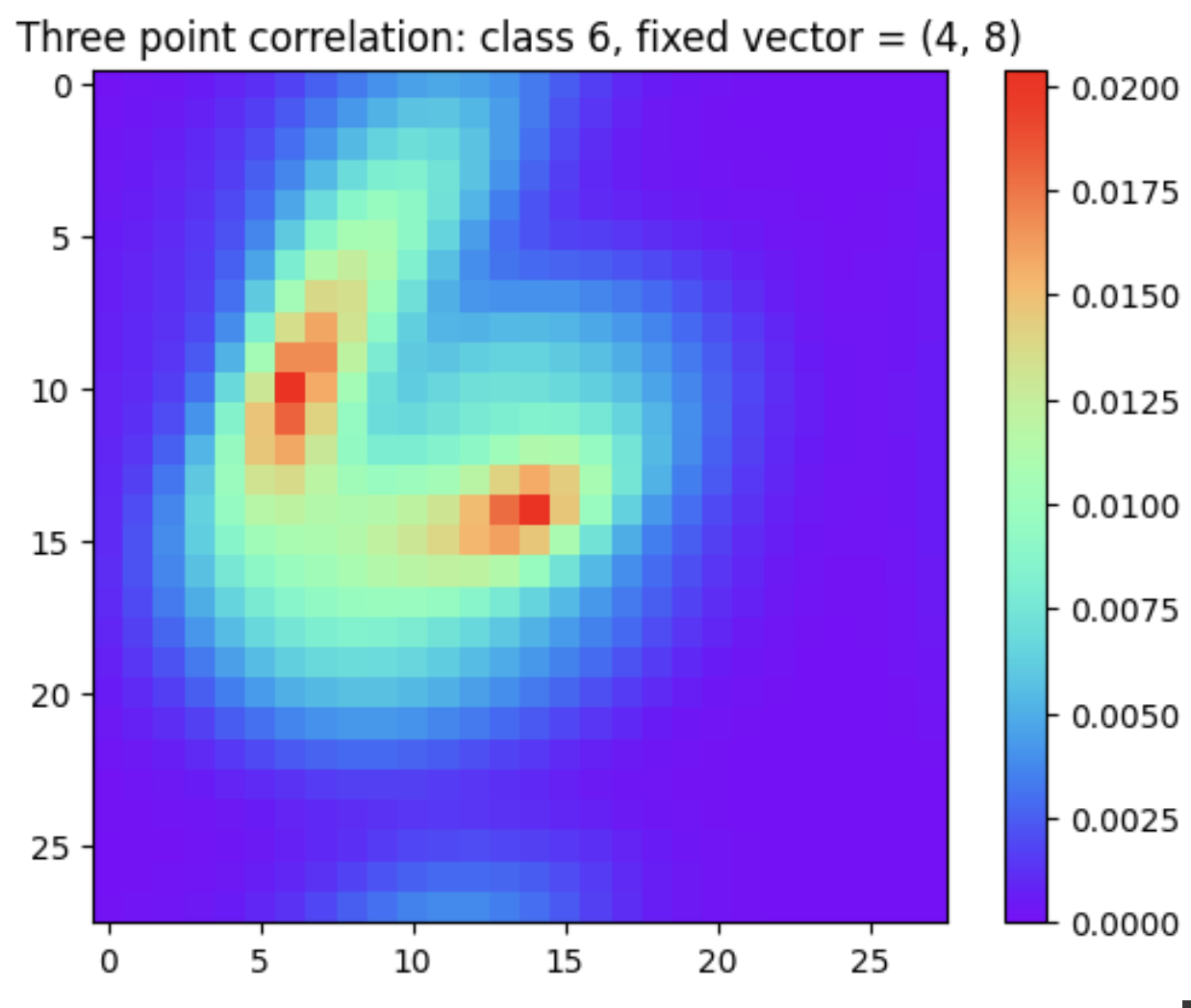}\par
       \includegraphics[width=\linewidth]{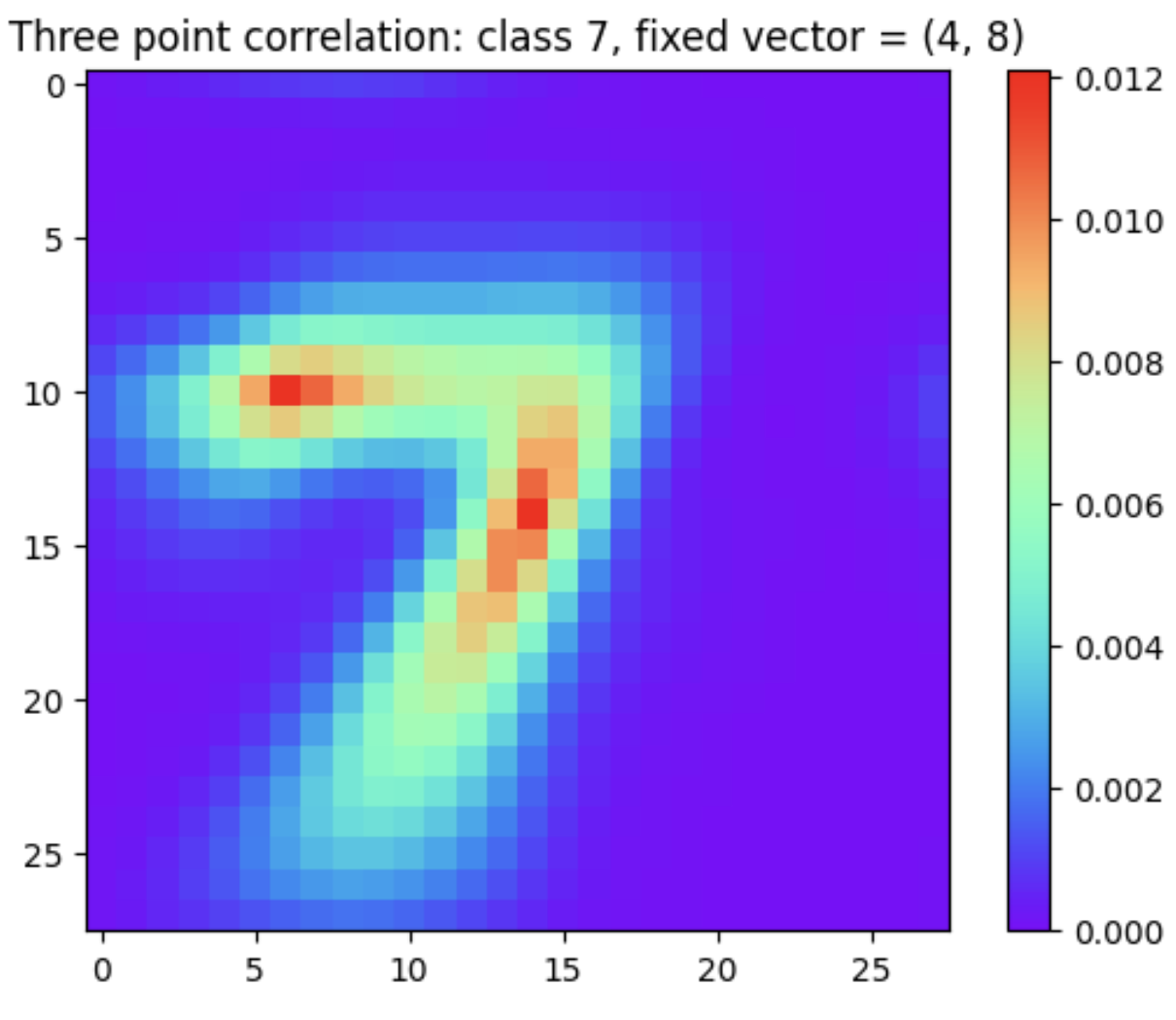}\par
       \includegraphics[width=\linewidth]{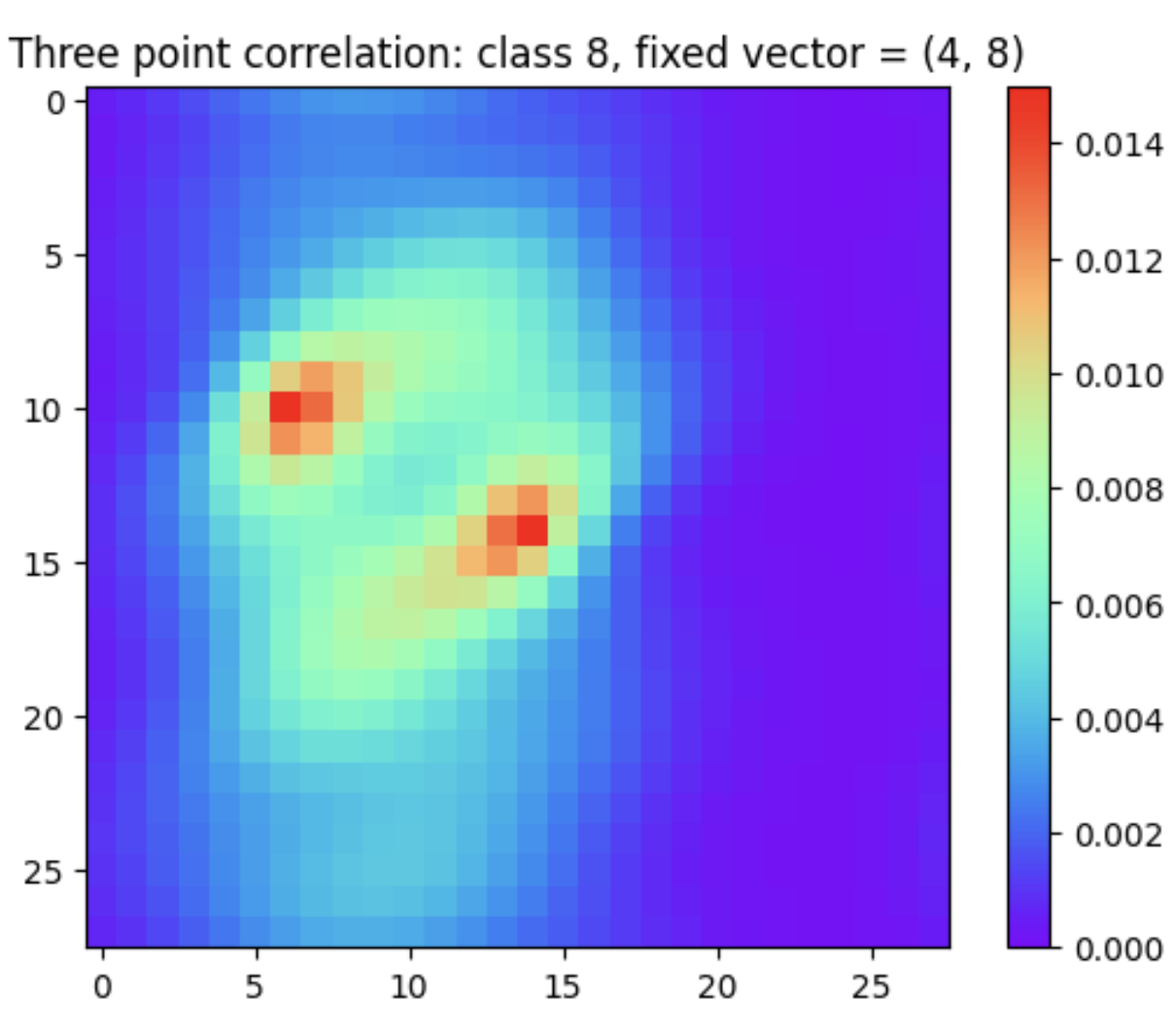}\par
          \includegraphics[width=\linewidth]{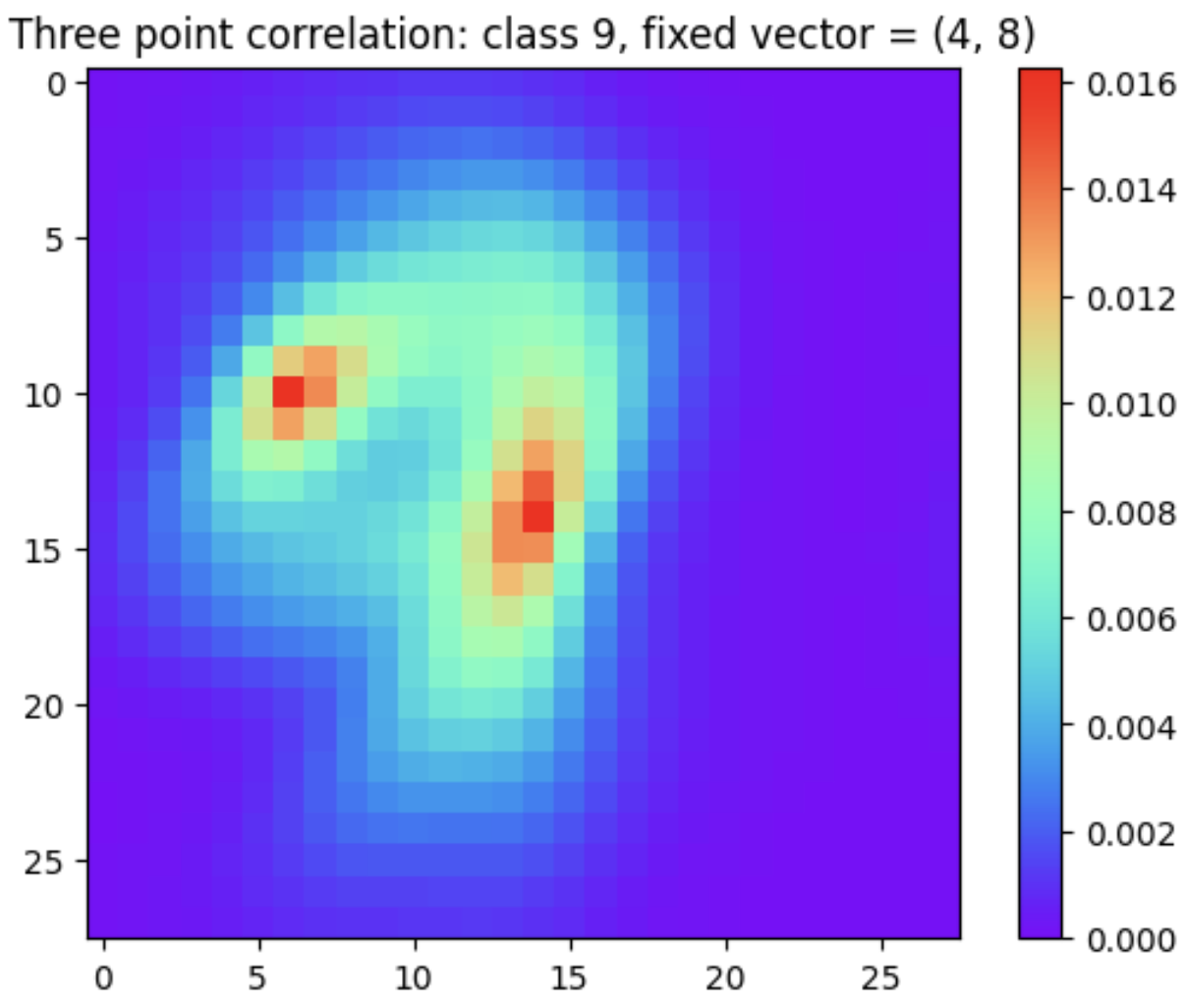}\par
    \end{multicols}

\caption{3-Point Probability Plots.  Scales on Right Shade from Lower Probabilities (Bottom) to Higher Probabilities (Top).}
\label{3-point-plots}
\end{figure*}

  Work in collaboration with Stephan Wojtowytsch\footnote{\url{https://www.mathematics.pitt.edu/people/stephan-wojtowytsch}.}
  examined the labelled numerals  in the MNIST dataset and shows that  $3$-point correlation functions are  able to begin to distinguish, say, sevens from fours. Compare the 2-point plots (figure~\ref{2-point-plots}) with the 3-point plots (figure~\ref{3-point-plots}).

To yield the results displayed in figures~\ref{2-point-plots} and \ref{3-point-plots}, the MNIST data were changed from grayscale to black and white, with white pixels given the value  1 and black pixels the value  0.  Each image is a square of $28^2$ pixels labelled by $x_{i,j}$.  We first seek  the probability that two pixels values $[x_{i,j}]$ and $[x_{i + \mbox{\footnotesize{shift}}_x}, x_{j + \mbox{\footnotesize{shift}}_y}]$ are both white (= 1) for some fixed $\mbox{{shift}} = (\mbox{\footnotesize{shift}}_x, \mbox{\footnotesize{shift}}_y)$.\footnote{``[$\cdot$]''= value of pixel [$\cdot$] $\in \{0,1\}$.} This (2-point) probability is given by:
\begin{equation}\begin{split}
Pr([x_{i,j}] & =    [x_{i + \mbox{\footnotesize{shift}}_x}, x_{j + \mbox{\footnotesize{shift}}_y}]) \\
&  =  \frac{1}{\mbox{\footnotesize{\#Images in {\bf x}}}}\sum_{{\bf x} }\frac{1}{28^2}\sum_{i =1}^{28} \sum_{j=1}^{28}C_{i,j}({\bf x}) = 1 \end{split}
\end{equation}
where
\begin{equation}C_{i,j}({\bf x})= \begin{cases} = 1, \mbox{if} \ [x_{i,j}] \cdot [x_{i + \mbox{\footnotesize{shift}}_x, j + \mbox{\footnotesize{shift}}_y}] = 1\\
					= 0, \mbox{otherwise}.\end{cases}\end{equation} 
For three point correlations we replace $C_{i,j}({\bf x})$ with $C_{i,j,{\bf{k}}}({\bf x})$ defined  (for ${\bf k} = (l,m)$---a given shift from pixel $x_{i,j}$ in the image) as follows:
\begin{equation}
C_{i,j,{\bf{k}}} = \begin{cases} = 1, \mbox{if} \ [x_{i,j}] \cdot [x_{i + \mbox{\footnotesize{shift}}_x, j + \mbox{\footnotesize{shift}}_y}] \cdot [x_{i + l, j + m}] = 1\\
= 0, \mbox{otherwise}.
\end{cases}
\end{equation}

\noindent This yields the $3$-point correlations displayed in  Figure~\ref{3-point-plots} for each class of numerals with the shift from pixel $x_{i,j}$, set to ${\bf k}= (l,m) = (4,8)$. It is evident from Figures~\ref{2-point-plots} and \ref{3-point-plots} that $3$-point correlations allow for distinctly better discrimination between the numerals in the MNIST dataset.  This, and other plots implementing the same formula (with different shift pixels ${\bf k}$), provide evidence that higher order correlation functions of the kind discussed briefly in section~\ref{cf_method}, are sufficient for distinguishing distinct classes of numerals in the  dataset. It is reasonable, we believe, to expect similar results from the other datasets mentioned listed in section~\ref{dataset}. 

We think the best way to interpret these plots is to consider them as progressively better approximating representative volume elements (RVEs) for the distinct classes of numerals. The analogy with other datasets being RVEs that distinguish, say, the class of dogs from the class of cats. Furthermore, we suggest a connection\footnote{We intend to elaborate on this connection in future work.} between the RVE concept in materials science and   the ``Information Bottleneck Method'' first suggested in \cite{bottle_tishby_bialek}. The Information Bottleneck Method (IBM) aims to find the ``relevant'' information, a ``sufficient statistic,'' in one set of variables (pixel inputs in the current context) for predicting values of some other variable (image labels in the current context). This method is typically suggested/applied in  contexts where much of the information in the former is irrelevant for the latter.\footnote{We purposefully cite the early Information Bottleneck paper (\cite{bottle_tishby_bialek}) and not later applications of  it to deep learning, because there are debates about whether some of these further applications and elaborations are correct. See \cite{IBM_deep_learning} for details.}
 We think that DNNs, during training via SGD, are implementing this strategy. That is, DNNs learn to discard most of the (irrelevant) information in the raw pixelated (input) format and to retain only that information that is relevant for image classification. This retained information constitutes the appropriate RVEs for the different classes within datasets.

Figures~\ref{2-point-plots} and \ref{3-point-plots} provide evidence that determining $N$-point correlation functions for $N>2$, does allow for  discriminating between different classes in (the MNIST) dataset. That is, it answers the first question posed at the beginning of this section, \ref{q1}. Nevertheless, the second question remains: 
\begin{enumerate}[label=Q\arabic*]
\setcounter{enumi}{1}
\item \textit{Can one show} that, as a matter of fact, in image recognition (and other tasks) DNNs are indeed finding such higher order ($N>2$) correlations.
\end{enumerate} How are these correlation functions being realized?  The next section provides some evidence that, in fact, DNNs are finding such correlations and offers a suggestion of how (at least theoretically) they are able to do so.

\subsection{Learning Higher Order Correlations \label{beyond_2pt}}

A recent paper entitled ``Neural Networks Trained with SGD Learn Distributions of Increasing Complexity'' \cite{Refinetti_Learning_distrib} proposes what the authors call the \textit{Distributional Simplicity Bias}.

\begin{itemize}
\item[(DSB):]  ``A parametric model trained on a classification task using SGD discriminates its inputs using increasingly higher-order input statistics as training progresses.'' \cite[p. 2]{Refinetti_Learning_distrib} \end{itemize}

\noindent In more detail,  the DSB  	
\begin{quote}
\textit{shifts the focus from characterising the function that the network learns, to identifying the features of the training data that influence the network}. We conjecture that any parametric model, trained on a classification task using SGD, is initially biased towards exploiting lower-order input statistics to classify its inputs. As training progresses, the network then takes increasingly higher-order statistics of the inputs into account. \cite[p. 1; Emphasis added]{Refinetti_Learning_distrib}
\end{quote}

\noindent This principle,   supports the connections made in the previous section between the empirical investigations of Levi and Oz  and those of Martin and Mahoney. We feel the DSB succinctly supports the main thesis of this paper. 

\begin{figure}[h]\centering
\includegraphics[height=.35\textheight, angle=0, trim=.9in 8.7in 6.33in .9in,clip]{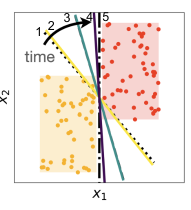}
\caption{\label{rect_data}  Rectangular Data and Decision Boundaries \cite[p. 4]{Refinetti_Learning_distrib}}
\end{figure}

They motivate the principle by considering a simple/toy model of a single perceptron (a simple linear classifier) that is trained to distinguish between two types of data points that reside in two distinct rectangles in a plane.  See figure~\ref{rect_data}. 
The data ${\bf x} = (x^i)_{i\leq D}$ are split into the two equally probable classes and are given labels $y_i$ with $i = \pm 1$.\footnote{In the figure these labels are yellow and red.}  As one can see, the optimal decision boundary between the two classes is a line (black dot-dashed, labeled ``5'') parallel to the $x_2$-axis that they call the ``oracle.'' \cite[p. 4]{Refinetti_Learning_distrib}. For details of this example, see the appendix \ref{appendix}. 

Here we  discuss briefly the stages of learning that the single perceptron model traverses as it is trained on the data in figure~\ref{rect_data}. The perceptron's equation yields the output \((\hat{y} \in \{\mbox{yellow}, \mbox{red}\}) \), given the input \[\lambda \equiv w_ix^i/\sqrt{D}.\] $\sigma$ is the activation function (in this case the sigmoid function) and  \[w = (w_i)\in \mathbb{R}\]  is the weight vector. During training (on a test set) the penalty for a wrong classification is the square loss: \[l(\lambda,y)  = (\sigma(\lambda) - y)^2.\]  As training progresses under SGD the form of the weight vector, $w$, evolves and leads to a clockwise rotation of the classifier line as shown in  figure~\ref{rect_data}.

In order to track this evolution Refinetti, et.\ al, expand the evolution equation for  the activation function $\sigma$ in a Taylor series around $\lambda = 0$. \cite[pp. 4, 14--15]{Refinetti_Learning_distrib}  This  allows them to look at the \textit{zeroth order} of the weight vector (first term in the series), the \textit{first order} (second term in the series), etc. 

\begin{itemize}\item They find that the zeroth order term is just the difference between the means of each data class. As a result of ``learning'' this information, the decision boundary rotates from the results of early training  from random initialization for  the weight vector $w$ (dotted black line, labeled ``1''),   to the ``naive'' classifier  (yellow line, labeled ``2'') which is determined by the equation at zeroth order.\footnote{This is our best attempt to interpret what is said in the figure caption and in the text. There are several inconsistencies.} This zeroth order depends on 1-point correlations (means) and the yellow classifier line splits these means  of the two rectangular classes.

\item At first order (second term in the series expansion) the perceptron rotates the classifier  to Fisher's Linear Discriminant (green line, labeled ``3'').  This term (and associated classifier line) maximizes the ratio  between the means of the two classes (their centers of mass) to the variance within each  class. That is, it maximizes the separation of the means while minimizing the variances or spreads in each class.\footnote{The discriminant has the form 
\(\frac{(\mu_2-\mu_1)^2}{\sigma_2^2 + \sigma_1^2}.\) See \cite[pp. 186--189]{Bishop_pattern_ML} for a clear explanation.} At this stage, the perceptron updating has taken the both the means and variances---the Gaussian statistics in to consideration.

\item At second order (third term in the Taylor expansion of $\sigma(\lambda)$) there is no contribution to the flow of the gradient  of $\sigma$. This is due to symmetry between the two classes of data. 

\item At third order, non-Gaussian statistics \textit{are} relevant and begin to lead to further improvements under SGD. This rotates the classifier line (now purple, labeled ``4'') even closer to the optimal classifier---the oracle.
\end{itemize}

This single perceptron model is an extremely  simple toy model.  It allows one to explicitly demonstrate how the consideration of higher order correlations can improve classification.  But, of course, this model is not really learning increasingly complex functions. As Refinetti, et al., note ``its decision boundary remains a straight line.'' \cite[p. 5]{Refinetti_Learning_distrib} Nevertheless, as noted, the \textit{direction} of the weight vector $w_i$, 
\begin{quote}
and hence its decision boundary, first only depends on the means of each class, \dots, then on their mean and covariance, \ldots, and finally also on higher-order cumulants, \ldots, yielding increasingly accurate predictors.  \cite[p. 5]{Refinetti_Learning_distrib} \end{quote}

While Refinetti, et al., do not make the connection, this procedure strongly resembles strategies for systematically extracting corrections to non-interacting Gaussian fields in Quantum Field theory. The so-called ``$\epsilon$-expansion'' aims to determine $N$-point \textit{non-Gaussian} correlations (Green's functions) via a perturbation expansion around a zeroth order Gaussian field.\footnote{See \cite[Section 4]{wilson-rg} and \cite[Chapter 12]{goldenfeld.book} for the details.}. Both procedures begin with Gaussians in the early phases of construction, and aim to find higher order correlation functions that yield more accurate corrections.  Importantly, in the current context, the ``expansion'' parameter in the Taylor series results from the \textit{SGD dynamics of updating during training} and shows how higher order statistics ($N$-point correlations for $N>2$) are important for classification.


As further evidence for their ``distribution centric'' (i.e., dataset centered) point of view, these authors train neural networks with different architectures on various approximate ``clones'' of the CIFAR dataset. \cite[p. 6]{Refinetti_Learning_distrib} These clones are designed to have the same mean and covariance as the images in CIFAR, but also differ  by progressively including higher order cumulants. \cite[pp. 6--8]{Refinetti_Learning_distrib} The results, while somewhat preliminary, provide further evidence that the neural networks are learning distributions of increasing complexity \textit{after having learned} the first and second order  correlational statistics.

Our goal in this section has been to provide some evidence in favor of the hypothesis that DNNs are successful at certain tasks (specifically, but not exclusively, image recognition tasks) because they  implement something like the correlation function methodology described in section~\ref{cf_method}. This evidence is reflected in arguments for what Refinetti et al.\ call the ``distributional simplicity bias.'' It explicitly appeals to  the stochastic gradient descent algorithm (SGD)  and aims to show  that the  algorithm is driven by progressively examining higher order dataset statistics. 

As noted, the idea here is similar to those developed in the context of quantum field theoretic perturbative calculations of $N$-point correlation (Green's) functions.  The idea there being that having $N$-point correlations with $N\to\infty$ enables the (re)construction of a quantum field. In analogy, we believe that being  able to determine $N$-point correlation functions of image datasets for $N>2$,  will provide  information sufficient for the successes of DNNs on image recognition tasks. As argued in section~\ref{cf_method} one way to conceptualize this is the following:  DNNs, in determining higher order correlation functions, are finding statistical representatives (RVEs) that uniquely distinguish   classes in a given dataset.

\section{Conclusion: The Importance of Worldly Structure \label{conclusion}}

We would like to stress that understanding the successes of DNNs on various tasks requires a focus on facts about the world. These worldly facts are responsible for robust  statistical properties that are present in the datasets upon which the DNNs are trained. The workings of DNNs will remain obscure and opaque if one only looks ``under the hood'' of the DNNs themselves and ignores these robust, universal, features of the datasets. Furthermore,  we believe that these statistical features present in the datasets  are  what drive the dynamics of  weight updating.   

\subsection{Connections with Multiscale Modeling}
In addition, we have argued that DNNs are actually implementing a well-understood methodology (important for any field that aims to explain continuum scale behavior of complex systems) that privileges correlational structures at mesoscales.\cite{middleway}  In the context of image recognition, the lowest (fundamental) scale corresponds to features of individual pixels such as their luminance, their weighted color (RGB) values, etc.  In analogy with many-body physical systems, the ``continuum'' scale behavior of images is their (correct) label---whether the image is that of a dog, a cat, etc.  As in the physical examples, the most important features or quantities for identifying the images as members of a given class are  functions that represent correlational structures in the images at mesoscales.  And, just as in the physical examples, these correlations  are hidden at the smallest pixel scale.  The discovery of these correlational features allows for successful identification of an image as a member of a specific class. 

\subsection{SLT vs.\ Deep Learning}
Our focus on correlational structures in datasets also highlights differences between understanding contemporary deep learning and the theoretical perspective of statistical learning theory.  As we noted, DNNs can successfully generalize to test sets upon training, despite commonly employing more parameters than the data points on which they are trained.  This is contrary to apparent implications of  SLT and, for that matter, conventional informal statistical wisdom about overfitting.  We suggested that part of the reason for this is a mismatch between the assumptions made in SLT and the empirical features of the data on which they are trained. 
SLT provides worst case bounds on expected performance on a test set without any  assumptions concerning the probability distributions characterizing the data.
However, as we have emphasized, real world data consisting of images is governed by very specific kinds of  probability distributions. DNNs, we suggest, operate so as to learn certain features of these probability distributions, particularly those having to do with higher order correlations. By contrast, for arbitrary probability distributions, higher order correlation functions may not exist or may be uninformative.  Moreover,  in constructing and training DNNs that classify well, what matters is not worse case behavior but something more like typical or attainable behavior. 

Another way of expressing this point is to note that, although the collections of images on which DNNs are trained have very high dimensionality, there is a great deal of evidence  (as one might expect) that the effective number of dimensions that the DNNs employ in classifying data is many orders of magnitude smaller.\footnote{This is sometimes described as the ``submanifold hypothesis,'' according to which the high dimensional manifold associated with the raw images contains a much smaller dimensional submanifold that contains the information relevant to successful classification. \cite{dimension_blessing}  It is worth noting that  the submanifold hypothesis can be correct even though the functions the DNNs implement are, by some standard, non-simple. Even if the effective dimension is small, it is entirely possible that the functions defined on that submanifold have a complicated structure.  This is also relevant to the Information Bottleneck mentioned earlier in section~\ref{npcf}.} This reflects the fact that there is considerable redundancy in real life images when these are viewed at the level of individual pixels  and the task is one of sorting them into rather coarse-grained categories (``dog'' vs.\  ``cat'').  
The presence of scale invariances and power law distributions in the characterization of images, detailed above, is one facet of such redundancy. Note again, that there is nothing a priori about this---one can certainly imagine a collection of pixels that does not have this kind of redundant structure. There, the exact luminance levels of pixels in comparison with other arbitrarily chosen pixels might be critical for how the collection is classified.
Presumably, however, collections of pixels without such redundant structure would not be recognizable by us as images of anything---they would look like noise. As Ruderman's work shows, images in the sense of scenes composed of  objects recognizable by us, have very different structures. When we train a DNN to classify in accordance with the classifications we make, we train it to pay attention to these structures. 
\subsection{Smoothness\label{smooth}}
Another observation: As discussed in section~\ref{obj_scale}, real images have the feature that pixels that are close by in physical distance are generally similar in luminance. One can think of this as a kind of smoothness condition---pixel luminances do not generally vary wildly over short distances.\footnote{Recall Ruderman's ``difference function'', equation~(\ref{diff_fun}). Again,  one can imagine a collection of pixels that does not have this feature, but it would not look like an ordinary image.} This smoothness condition is another ``worldly condition'' that characterizes real life images.  

It is tempting to make the following connection regarding smoothness. As Martin and Mahoney \cite{Martin_Mahoney_rmt} (and others) have shown, stochastic gradient descent implements ``self-regularization.'' That is, it selects functions that have low norm.\footnote{In either the $l_1$ or $l_2$ norm, these, roughly, are  those with ``small'' coefficients.} Such functions are also relatively smooth---they do not, for the most part,  exhibit large changes over small distances except at object boundaries. In this respect they are well matched to the smoothness condition satisfied by real life images.  One may then conjecture that such self-regularization selects for functions that track these features,  thereby helping to explain why SGD leads to results that ``work'' for images. If this is correct, one would expect that DNNs trained with SGD would work less well with structures that do not satisfy smoothness conditions and this in fact is what is found. See the brief discussion of tabular data below. \cite{tree_tabular,survey_DNN_tabular,simplicity_bias}

\subsection{More on What it Means to say that the Data Matters}

We have argued throughout this paper that the structure of the data matters to whether successful generalization by DNNs occurs. DNNs, as currently configured, have capabilities that learn specific features present in images but not necessarily in other data---features that support generalization to new data. One might wonder, however, about how exactly to understand ``data matters.''  In particular, as noted in section~\ref{intro}, one might contend that at least some of the data features we have identified as crucial to DNN learning could equally well be conceptualized as restrictions on the function class (the effective capacity) of the DNN. For example, if the architecture of a DNN implements a preference for low-norm functions that fit well with image data, then (it might be argued)  this might just as well be regarded as an ``effective'' restriction on the function class that the DNN is able to implement. Of course, this restriction need not be explicitly put in by a human modeler and may instead arise``implicity'' from, e.g., the use of use of SGD and perhaps other features of the network, but we can, nonetheless, think of it as a restriction on the function class.

While we agree that such a reconceptualization is sometimes possible\footnote{We say ``sometimes'' because there are  some data based features that affect learning that are not plausibly viewed as restrictions on the function class that a DNN is able to implement, even on a very generous conception of ``function class.'' These include ``fast rate'' restrictions as in \cite{fast_convergence_kim}. Simplifying greatly, these results have to do with the extent to which the  majority of the data happens to be close to the decision boundary in a classification problem. Of course such closeness negatively affects learning. Whether it is present or not is a feature of the data distribution which can vary independently of the set of functions the DNN is able to implement. As it happens, natural images tend to have favorable decision boundaries-- it is usually not uncertain or ambiguous whether an image is a dog or a cat.  This is another data-dependent feature that contributes to their learnability.}
, we do not think it undermines anything we say in this paper.  If our argument is correct, the structural features of data to which we draw attention will affect generalizability regardless of whether or not they are understood as a restriction on capacity. 

As an illustration, consider  Belkin, et al. \cite{belkin_bias_var_tradeoff}, who show that certain DNNs have the capacity to fully memorize data with random labels during training, although of course there was no successful generalization to the test set. The same DNNs do, nevertheless, learn to successfully generalize on ``real'' data. 
The capacity of the DNNs, including any implicit restrictions associated with, e.g., a preference for low-norm functions,  is the same across the two kinds of cases. In the cases involving random data, the DNNs presumably try to find fitting functions that are as low norm as possible, just as they do with real data.  The fitting functions may be as low norm as possible given the random nature of the data, but they still presumably will be relatively high norm in comparison with the functions used to fit real data. For real data, the norms of the fitting functions will be much lower and this will make a difference for successful generalization. In other words, regardless of where we locate the preference for low norm functions (in the capacity of the function class or elsewhere), the consequences for successful generalization will still depend on the nature of the data.

Another way of bringing out the significance of the point that the data matters, is this: If it is correct, as we have argued, that DNNs succeed in learning to correctly classify images because they track certain specific features of those images, one would expect DNNs to be less successful on data that lack these features. As remarked above, there is considerable empirical support for this claim.  A substantial amount of recent research has focused on the use of DNNs on tabular data. Tables lack many of the features that we have emphasized in connection with images---they lack the strong local correlations and smooth within-object variation present in images, as well the invariances that characterize them. Unsurprisingly, DNNs perform considerably less well on tables and less well than alternative approaches that embody ``inductive biases'' that are more appropriate to tables such as tree-based models-- for discussion, see e.g. \cite{tree_tabular,survey_DNN_tabular} 

Still another way of bringing out the significance of  our claims about the data mattering is in terms of the following contrast: A common philosophical assumption (and one that is shared by at least some statisticians)\footnote{Sober's \cite{sober_reconstructing} discussions of the role of simplicity considerations in evolutionary biology are an important exception. Like us, Sober holds that  when hypothesis preferences based on simplicity work, this is because they encode empirical assumptions about the domain to which they are applied.}  is that there is a domain- or data-independent ordering of hypotheses that provides a normative basis for selecting hypotheses that not only fit some known body of data (the training set in the context of DNNs) but are most likely to generalize well to new data (the test set in the context of DNNs). By ``data-independent'' we mean that the ordering is appropriate to use on any set of data, regardless of its structure.\footnote{Of course, someone holding this view can acknowledge that successful generalization may be harder from some kinds of data as opposed to others. But one can still be committed to data independence in the sense that there is a fixed  ordering of  the possible hypotheses one should employ for purposes of generalization, regardless of what the data is like---linear hypotheses  are still the best first choice even if they end up being unsuccessful.}   Sometimes this is put in terms of a data-independent notion of simplicity. Philosophers who claim that linear functions are simpler than quadratic relations and hence that the former should be privileged in generalization hold such a view, as do those who claim that hypotheses that contain fewer free parameters are automatically preferable from the point of view of generalization to new data.  Popper's claim that alternative hypotheses can be ordered in terms of their degree of falsifiability and that, other things being equal, when generalizing one should try the most falsifiable hypothesis first is yet another example.  Finally, measures of capacity employed in statistical learning theory, like those based on shattering, are also often understood to have this feature of data independence.   
 
          Our claims about the structure of the data mattering can be understood as the denial of the position just described. In other words, we claim that any ordering of hypotheses that will track successful generalization cannot be universal or domain-independent in this sense. The example of DNNs illustrates this point.  A preference for relatively smooth or low norm functions of the sort that appears to be realized in DNNs using SGD fits with, and is appropriate for, generalization over data consisting of visual images because it matches features that, as an empirical matter,  are present in those images.  However, such a preference  appears to be inappropriate for other sorts of data such as tabular data, as we observed above.


        We also note in this connection that the notion of simplicity, so often invoked in these contexts, is rather flexible and, by itself, not very illuminating. One can claim, as some researchers do (e.g. Belkin et al. \cite{belkin_bias_var_tradeoff}), that low norm and other functions that DNNs find are ``simple.'' But, even if this is correct by some standard of simplicity,  it does not, by itself,  explain why such functions work for purposes of generalization. Instead, it is the fit between such functions and the relevant features of images that matters for generalization. This suggests that simplicity claims, when useful in guiding generalization, are useful  to the extent that they encode relevant empirical features of the domain of interest that support generalization. 

      We take the observations in this section to illustrate how the problem of understanding successful generalization in DNNs  has important connections with standard questions and assumptions in philosophy of science that concern how ``inductive'' reasoning and learning from evidence work, and whether data-independent accounts of learning and generalization are likely to be successful. There are also important issues we have only gestured at: These include, what exactly is involved in ``overfitting,'' and when and why it is deleterious. In general, the success of DNNs suggests that considerable rethinking of these issues may be required.

\newpage
\appendix
\section{Appendix: Details of the Toy Perceptron Model of Section~\label{appendix}}

The perceptron's output is given by:
\begin{equation}
\hat{y} = \sigma{\lambda}, \quad \mbox{for}\quad  \lambda \equiv w_ix^i/\sqrt{D},
\end{equation}
with the weight vector $w_i$ and a nonlinear (sigmoid) activation function $\sigma$.\footnote{Superscripts are indices for inputs and subscripts are indices for weights.} The penalty during training is given by the square loss: \[l(\lambda,y)  = (\sigma(\lambda) - y)^2.\] The perceptron is initialized with weights drawn from a Gaussian  with variance 1 and then trained on a test set. They examine how the  weight vector \textit{evolves during SGD} with the given loss function. That is, they study the gradient flow:

\begin{equation}\dot{w_i} = \eta \mathbb{E} (\sigma(\lambda - y)\sigma'(\lambda)x^i,
\label{gf}\end{equation}
with fixed learning rate $\eta>0$.  Here $\mathbb{E}$ is an average over the data distribution. \cite[p. 4]{Refinetti_Learning_distrib}. They do a Taylor expansion of the activation function, $\sigma(\lambda)$, around $\lambda = 0$:
\[\sigma(\lambda) = \sum_{k=0}^\infty \beta_k\lambda^k.\]  This gives the derivative\footnote{We have corrected a typo in this equation.} \[\sigma'(\lambda)\sum_{k=0}^\infty \tilde{\beta}_{k+1}\lambda^{k-1},\] 
with \(\tilde{\beta_k} = k\beta_k = k\sigma^{(k)}(\lambda)|_{\lambda = 0}.\)

When entered into the gradient equation~(\ref{gf}) this yields:
\begin{equation}\label{taylor}
\frac{1}{\eta}\dot{w_i} = \sum_{k=0}^\infty\mathbb{E}\lambda^kx^i(\gamma_k - \tilde{\beta}_{k+1}y),
\end{equation}
with $\tilde{\beta}_{k}$ and $\gamma_k$ constants from the Taylor expansion.
This allows them to look at the \textit{zeroth order} through \textit{third order} components---the first four terms in the expansion (\ref{taylor}) of the gradient flow equation~(\ref{gf}) of the perceptron weight as it is trained.\footnote{This model just focuses on the evolution of the perceptron's  weight $w_i$ and ignores the bias (which is fixed).}

They show that at \textit{zeroth} order the ``weight converges \textit{in direction}'' as
 \begin{equation}w_i^{0} \propto m^i \equiv \kappa_+^i - \kappa_{-}^i.\end{equation}
 \textit{This  is the difference between the means of each class corresponding to $1$-point correlations or statistics}, \[\kappa_\pm^i = \mathbb{E}_\pm x^i.\]  
 The idea here is that under training of this simple model, the decision boundary between the two classes rotates in the direction of the (curved) arrow in figure~\ref{rect_data}.  At this zeroth order, the decision boundary moves to a line (yellow, labeled ``2'') splitting the means of the two rectangular classes.

At \textit{first} order, the gradient of $w_i$ depends upon the \textit{second moments corresponding to $2$-point correlations} in the (data) inputs taking into consideration the difference between the ``within class'' variances and the variances ``between classes.''  This further rotates the decision boundary to the right and yields the classifier $w_i^{(1)}$, known in statistics as ``Fisher's linear discriminant.''\footnote{See \cite[pp. 186--189]{Bishop_pattern_ML} for a clear explanation.}   \cite[p. 4]{Refinetti_Learning_distrib} In figure~\ref{rect_data}, this is the green line (labeled ``3'').\footnote{We believe that the text here misidentifies the Fisher  discriminant as a ``dashed black line.''}

Next, the \textit{second} order term in the expansion contributes nothing to the gradient flow as it would involve the third moment which in this case is equal to zero because of  symmetry in the data. As a result, the second order term in the perturbation expansion of the gradient flow does not improve upon the first order term.

    Finally, the \textit{third} order  involves the fourth moments: $\kappa^{ijkl}$.  As with the second moments, they decompose the fourth moment into a ``between-class'' fourth moment and a ``within-class'' fourth moment. Without going into the details, this allows them to express the fourth moments in terms of contributions from a ``within-class'' fourth order \textit{cumulant} and contributions from the mean and second moments. At this order the expansion finally  takes into consideration some \textit{non-Gaussian} correlational information that is present in the dataset.\footnote{Cumulants at a given order are non-zero if there are genuine correlations that cannot be accounted for by correlational structure at lower orders. They are sometimes (in the literature on QFT) called ``connected correlation functions.'' For instance, second order cumulants have the form $\langle XY \rangle - \langle X\rangle\langle Y \rangle$ which subtracts trivial contributions from the means. If $X$ and $Y$ are independent, the cumulant is zero.}\newpage  
    \noindent This correlational information rotates the decision boundary to the purple line (labeled ``4'') which is even closer to the ``oracle.''

\newpage


\end{document}